\documentclass[aps,prb,10pt,twocolumn]{revtex4-2}
\usepackage{graphicx,overpic}
\usepackage{amssymb,amsmath,mathtools,textcomp}
\usepackage{bbm,enumerate}
\usepackage{subcaption}
\usepackage[newcommands]{ragged2e}
\usepackage{tabularx,colortbl}
\usepackage{array}
\usepackage{multirow}
\usepackage{comment}
\usepackage{ bbold }

\newcommand{\lio}{Li$_2$IrO$_3$}

\newcommand{\rucl}{\mbox{$\alpha$-RuCl$_3$}}

\newcommand{\w}{\omega}

\newcommand{\Bcrit}{B_{\rm c}}

\newcommand{\hcrit}{h_{\rm c}}

\newcommand{\ha}{\vec{B}\!\parallel \!\hat{a}}
\newcommand{\hb}{\vec{B}\!\parallel \!\hat{b}}
\newcommand{\hc}{\vec{B}\!\parallel \!\hat{c}^*}
\newcommand{\hpc}{\vec{B}\!\perp \!\hat{c}^*}
\newcommand{\jeff}{j_{\rm eff}}
\newcommand{\hk}{Heisenberg-Kitaev }
\newcommand{\hkg}{HK$\Gamma$}

\newcommand{\dJ}[1]{\delta J_{#1}}
\newcommand{\dK}[1]{\delta K_{#1}}
\newcommand{\dG}[1]{\delta \Gamma_{#1}}
\newcommand{\kgp}{$\delta K/\delta\Gamma +$}
\newcommand{\kgm}{$\delta K/\delta\Gamma -$}

\newcommand{\translation}{$T$}
\newcommand{\inversion}[1]{$\mathcal{I}_{#1}$}
\newcommand{\spinflip}{$\mathcal{T}$}

\newcommand{\rotation}{$C_3$}
\newcommand{\flip}[1]{$C_{2#1}$}

\newcommand{\reflection}[1]{$C^*_{2#1}$}

\newcommand{\lllangle}{\langle\!\langle\!\langle}
\newcommand{\rrrangle}{\rangle\!\rangle\!\rangle}

\graphicspath{{./}{./figs/}}

\date{\today}

\begin{document}

\title{
Bond disorder in extended Heisenberg-Kitaev models: \\
Spin textures and in-gap states in the high-field regime
}

\author{Georgia Fragkopoulou}
\author{Matthias Vojta}
\affiliation{Institut f\"ur Theoretische Physik and W\"urzburg-Dresden Cluster of Excellence ct.qmat, Technische Universit\"at Dresden, 01062 Dresden, Germany}

\begin{abstract}
We study the effect of bond disorder in extended Heisenberg-Kitaev models on the honeycomb lattice, relevant for materials such as $\alpha$-RuCl$_3$, in the semiclassical limit using a combination of T-matrix and real-space spin-wave approaches.
Focusing on the regime of large applied magnetic field, we discuss two distinct but related disorder-induced phenomena, namely spin textures in the vicinity of isolated impurities and magnetic excitations below the bulk gap.
A finite impurity concentration smears the field-tuned phase transition and turns the isolated in-gap states into impurity bands. As a result, there is a large field regime above the bulk transition into the high-field phase where impurity-induced states fill the bulk spin gap. We illustrate the field dependence of these in-gap states for parameters relevant for $\alpha$-RuCl$_3$, and we connect our results to heat-transport and NMR data which indicated their presence.
\end{abstract}

\maketitle


\section{Introduction}

Mott-insulating magnets with strong spin-orbit coupling are a major research field in condensed-matter physics \cite{trebst2017,winter2017b,janssen19,takagi2019}. This has been partially triggered by Kitaev's construction \cite{kitaev2006} of a quantum spin liquid driven by bond-anisotropic exchange interactions on the honeycomb lattice, and by subsequent proposals \cite{jackeli2009,chaloupka2010,liu18,motome20} to realize Kitaev interactions in layered honeycomb magnets with $\jeff=1/2$ moments.

Among the candidate materials, iridates of the family $A_2$IrO$_3$ ($A=$Na, Li) and \rucl\ have received particular attention. However, these materials exhibit antiferromagnetic order at low temperatures, except for H$_3$LiIr$_2$O$_6$ which appears heavily disordered at a structural level \cite{kitagawa18,valenti18,knolle19}. The presence of antiferromagnetic order is attributed to magnetic interactions other than the Kitaev exchange, which tend to destroy the Kitaev spin liquid \cite{chaloupka2010,rau2014}. This order can be suppressed by applied magnetic fields, for instance at around 25~T for \mbox{$\alpha$-\lio} \cite{choi2019} and at 7--8~T for \rucl\ \cite{sears2015,johnson2015,leahy2017,baek2017,sears2017,wolter2017,zheng2017,hentrich2018}.
In {\rucl} the existence of a quantum spin-liquid phase in a narrow field range above the transition is suggested by a number of experimental results, such as an excitation continuum in neutron scattering \cite{banerjee2018,balz2019} and an approximately half-quantized thermal Hall conductivity \cite{kasahara2018b,yokoi20}. However, a definite consensus has not been reached, among others owing to significant sample-to-sample variations \cite{kasahara22,bruin22,zhang23,zhang24}.

Remarkably, also the asymptotic high-field phase of Kitaev magnets hosts interesting physics. Saturated magnetization at low temperature is only reached in the hypothetical high-field limit, and for generic directions of applied field the field-induced uniform magnetization does \emph{not} strictly point in field direction, but displays a significant transverse component \cite{janssen2017}. The high-field magnon excitations are gapped and can display non-trivial topological invariants \cite{mcclarty18,joshi18}.

For \rucl\ the field dependence of the bulk magnon gap has been mapped out \cite{baek2017,sears2017}, it reaches about 40~K at 15~T.
Therefore, it came as a surprise when experiments in {\rucl} indicated the presence of low-energy magnetic excitations at fields of 15~T and above: Heat transport measurements found a significant field dependence of the low-temperature thermal conductivity in this field range \cite{hentrich2020}, and NMR measurements showed a saturation of the temperature dependence of the spin relaxation rate at low temperatures \cite{baek2017,baek2020}.
A plausible interpretation of these phenomena is that they are driven by quenched disorder. Since the mechanism for dominant Kitaev exchange relies on details of the bond geometry \cite{jackeli2009,chaloupka2010}, non-magnetic crystalline defects likely have a stronger impact on nearby magnetic exchange couplings compared to more conventional non-Kitaev magnets. Disorder-induced low-energy excitations at high fields have also been detected in the Kitaev-candidate material Na$_2$Co$_2$TeO$_6$ \cite{xiang23}. These observations suggest that quenched disorder plays a significant role in Kitaev materials and raise questions about other little-understood features possibly related to disorder \cite{imamura_foot}.

Currently, there is limited concrete modeling of disorder-induced phenomena for this class of materials.
Several studies have addressed disorder physics in the Kitaev spin liquid \cite{willans2010,willans2011,zschocke15,knolle19,nasu2020,singhania2023}; others have studied general aspects of impurity-induced states in gapped quantum magnets without the intricacies of spin-orbit coupling \cite{utesov14,utesov2021,arlego2004,brenig2013}.
However, there is relatively little research on defects in extended \hk models relevant for {\rucl} and other Kitaev materials. Some papers have dealt with magnetic vacancies \cite{trousselet2011,andrade14,andrade20,li2023}, but systematic studies of bond disorder, in particular in applied magnetic fields, are lacking, with the exception of Ref.~\onlinecite{xiang23} specific to Na$_2$Co$_2$TeO$_6$.

This paper aims to address this gap. We study the physics of bond disorder in the relevant Heisenberg-Kitaev-Gamma (\hkg) models on the honeycomb lattice in the zero-temperature limit, with a focus on strong magnetic fields that drive the bulk system to the asymptotic high-field phase. We consider various types of isolated impurities and find cases where they induce spin textures in their vicinity and/or magnetic excitation modes inside the bulk gap. These results guide our proposal and analysis of impurity distributions at finite concentrations, which lead to low-energy excitations over a broad field range above the bulk critical field, as suggested by experimental data. Our results show that properties of the disorder-induced states are sensitive to the direction of applied magnetic field, and we provide concrete predictions for the anisotropic response of such states.

The remainder of the paper is organized as follows:
In Sec.~\ref{sec:model} we introduce the {\hkg} model and discuss general aspects of quenched disorder in Kitaev materials. Sec.~\ref{sec:swt} outlines the methodology of spin-wave calculations in the presence of disorder. Results for isolated impurities, including discussions on spin textures and their associated in-gap modes, are presented in Sec.~\ref{sec:1imp}. Sec.~\ref{sec:manyimp} shows results for finite impurity concentrations, particularly in relation to experiments on {\rucl}.
A summary of our results, together with suggestions for future experiments, closes the main part of the paper.
Finally, appendices provide a more in-depth analysis of various spin textures.


\section{Exchange models and bond disorder}
\label{sec:model}

\begin{figure}[tb]
\includegraphics{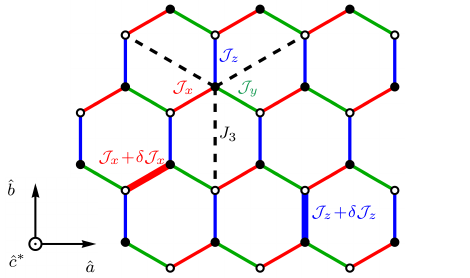}
\caption{
Honeycomb lattice with bond-dependent first-neighbor interactions $\mathcal{J}_{x,y,z}$ on red, green and blue bonds, respectively, and third-neighbor Heisenberg interactions $J_3$ (dashed lines); the $\mathcal{J}_{x,y,z}$ are $3\times 3$ matrices including Heisenberg, Kitaev and off-diagonal $\Gamma$ spin exchanges.
Nearest-neighbor bond defects $\delta\mathcal{J}_{x,y,z}$ are randomly placed in the lattice (bold lines).
}
\label{fig:latt}
\end{figure}

\subsection{Heisenberg-Kitaev-Gamma model}
\label{sec:hkgmodel}

The spin Hamiltonians proposed for {\rucl} and other Kitaev materials, often dubbed {\hkg} models, are extensions \cite{rau2014} of the honeycomb-lattice \hk model originally introduced in Ref.~\onlinecite{chaloupka2010}. Here, we will consider
\begin{equation}\label{Hamiltonian}
H=
\sum_{\langle ij\rangle_\gamma}
\vec{S}_i^{\,\mathrm{T}} \mathcal{J}_\gamma\, \vec{S}_j
+J_3 \sum_{\lllangle ij\rrrangle}  \vec S_{i} \cdot \vec S_{j}
-\vec h \cdot \sum_{i} \vec S_{i}.
\end{equation}
$\mathcal{J}_\gamma$ are the bond-dependent nearest-neighbor exchange matrices with
\begin{equation}
\mathcal{J}_x=\begin{pmatrix}
J+K & \Gamma' & \Gamma' \\
\Gamma' & J & \Gamma \\
\Gamma' & \Gamma & J
\end{pmatrix}
\end{equation}
and cyclic permutations, where $J$ and $K$ are the Heisenberg and Kitaev couplings, respectively, while $\Gamma$ and $\Gamma'$ parametrize symmetric off-diagonal terms. $J_3$ is a Heisenberg coupling between third-nearest-neighbor spins, and the uniform magnetic field is $\vec h \coloneqq g \mu_\mathrm{B} \vec{B}$, with $g$ the anisotropic effective $g$ tensor and $\mu_\mathrm{B}$ the Bohr magneton. We will study the model in the semiclassical limit, formally corresponding to large spin size $S$; the results apply semi-quantitatively also to the experimentally relevant case of $S=1/2$.

In Kitaev materials such as \rucl, crystallographic axes can be taken as $\hat{a}$ and $\hat{b}$ that are both located in the honeycomb plane of magnetic atoms, perpendicular and parallel to a Ru--Ru bond, respectively, and $\hat{c}^*$ perpendicular to the honeycomb plane, Fig.~\ref{fig:latt}. In spin space, using the frame of Eq.~\eqref{Hamiltonian}, they correspond to directions $\hat{a}\!\parallel\! [11\overline{2}]$, $\hat{b}\!\parallel\! [\overline{1}10]$ and $\hat{c}^*\!\parallel\! [111]$.

For the {\hkg} model \eqref{Hamiltonian}, different parameter sets have been proposed to describe \rucl, based on either ab-initio modeling or on fits to experimental data, and we refer the reader to Refs.~\onlinecite{janssen2017,maksimov2020,laurell2020} for an overview.
Guided by previous studies \cite{kim2016,winter2016,winter2017a,janssen2017,wangdong2017,winter2018,wolter2017} we employ parameters where $K<0$ and $\Gamma>0$ are the dominant couplings, while both $J<0$ and $J_3>0$ are small, mainly acting to stabilize the zigzag phase.
Most explicit numerical results in this paper are shown for $S=1/2$ and the parameter set $(J,K,\Gamma,\Gamma',J_3)=(-0.5,-5.0,2.5,0,0.5)\,\mathrm{meV}$. For the $g$ tensor we use $g_{ab}=2.5$ and $g_{c^*}=2.3$ for in- and out-of-plane fields respectively, as proposed in Ref.~\onlinecite{li2021}. While this combination leads to a somewhat larger in-plane anisotropy of critical fields than that experimentally measured \cite{lefrancois2023}, we expect our results be independent, on a qualitative level, of minor deviations in the exact parameters.

\subsection{Symmetries}
\label{sec:model:symmetries}

The strong spin-orbit coupling underlying Eq.~\eqref{Hamiltonian} breaks all continuous symmetries, leaving a distinct set of discrete symmetries. In the absence of quenched disorder, translation \translation\ and spatial inversion \inversion{} with respect to the center of any bond are symmetries of $H$ in Eq.~\eqref{Hamiltonian}.

Provided that $\vec{h}=0$, the Hamiltonian is also invariant under time reversal \spinflip, which is an anti-unitary transformation that acts on spins as $\vec{S}\rightarrow -\vec{S}$.

Furthermore, there are symmetries that involve simultaneous transformations of the spins and the lattice. One is a \rotation\ symmetry that rotates the lattice by $120^\circ$ about one of the lattice sites and the spins about the $[111]$ axis so that
$S^x \rightarrow S^y \rightarrow S^z \rightarrow S^x$.
Additionally, Eq.~\eqref{Hamiltonian} is also symmetric under rotation of spins such that $S^\alpha \rightarrow -S^\beta$, $S^\beta \rightarrow -S^\alpha$ and $S^\gamma\rightarrow -S^\gamma$ when the lattice is reflected with respect to the $\gamma$-bond. There are three symmetries of this type corresponding to the three different bonds that we label as \flip{x}, \flip{y}, \flip{z}.

The symmetries listed above are reduced when a magnetic field is applied. While \translation\ and \inversion{}\ still remain, \spinflip\ is always broken. The symmetries acting on both real and spin space are generically broken unless the field is applied in certain high-symmetry directions, when some or a combination of some can still be preserved.
For fields along the high-symmetry directions $\hat{a}$, $\hat{b}$, and $\hat{c}^*\!$, as defined in Sec.~\ref{sec:hkgmodel}, the following symmetries are preserved, as also summarized in the first row of Table~\ref{table:symmetries}.

(i) For a magnetic field $\ha$, the Hamiltonian is symmetric under the spin transformation $S^x \leftrightarrow S^y$ and reflection of the lattice along the $\hat{a}$ axis. We call this symmetry operation \reflection{z} and it is a combination of the fundamental symmetries of the model \inversion{}\spinflip\flip{z}. The same symmetry is present for all fields of the form $[11z]$ which lie on the plane perpendicular to the $z$-bond. There exist three of these planes corresponding to three symmetries \reflection{x}, \reflection{y}, \reflection{z}.

(ii) For $\hb$, which is parallel to the $z$-bond, \flip{z} remains a symmetry of the Hamiltonian. Equivalently, \flip{x} and \flip{y} are symmetries when the magnetic field is parallel to the x- and y-bond respectively.

(iii) The Hamiltonian with $\hc$ remains invariant under {\rotation} rotations. Additionally, since the magnetic field is perpendicular to all three bonds, it preserves the symmetries \reflection{x}, \reflection{y}, \reflection{z}.

\subsection{Field evolution of magnetic order}
\label{sec:model:ground_state}

Systems governed by the Hamiltonian \eqref{Hamiltonian} typically display symmetry-breaking long-range order at low temperatures. While a plethora of phases have been found as function of model parameters and magnetic field \cite{janssen19,janssen2016,rau2014,gotfryd2017,yilmaz2022}, we restrict our discussion to the physics encountered for the parameter set specified above which is believed to be relevant for \rucl.

\begin{figure}[bt]
\centering
\includegraphics[scale=1]{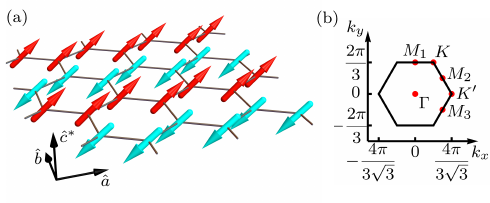}
\caption{
(a) One of the six zigzag ground states of Eq.~\eqref{Hamiltonian} for a parameter set relevant to \rucl.
(b) First Brillouin zone with high-symmetry points. A zigzag state perpendicular to the z bond as in (a) shows Bragg peaks at $M_1$.
}
\label{fig:zigzag}
\end{figure}

At $\vec{B}=0$, the ground state of \eqref{Hamiltonian} exhibits collinear single-$Q$ zigzag magnetic order, consisting of ferromagnetic chains being antiferromagnetically aligned to each other, as shown in Fig.~\ref{fig:zigzag}(a). The three distinct chain directions, together with global spin flip, yield a six-fold degeneracy of the ground state. The zigzag order is characterized by a Bragg peak at one of the three $M$ points of the Brillouin zone, Fig.~\ref{fig:zigzag}(b), depending on the direction of the zigzag chains. The ground-state spin direction depends of the ratio of $K$, $\Gamma$, and $\Gamma'$ \cite{chaloupka2016,hou2017,winter2017a,winter2018,kelley2018}. For the chosen parameters it is about $40^\circ$ out of plane; we note that experimentally measured moment directions differ if the $g$ tensor is anisotropic \cite{kelley2018}.

Adding a small magnetic field leads to canting of the zigzag order and, depending on the field direction, reduces the ground-state degeneracy. Specifically, for $\hb$ (or any field along a bond) the zigzag chains are oriented perpendicular to the field; while for $\ha$ (or any field perpendicular to a bond) the two zigzag orientations at $60^\circ$ with the field remain degenerate. Only for the special case of $\hc$ the three zigzag orientations remain degenerate: the \rotation\ symmetry is left intact by the field and then spontaneously broken by the ordered state. For a generic field direction, one zigzag orientation is uniquely selected.

At sufficiently high magnetic fields, the system reaches the asymptotic high-field phase. Generically, this displays a transverse magnetization which, however, vanishes for $\hc$ or $\hpc$ \cite{janssen2017}, which are the field directions of primary interest in this paper. For the parameters we use, Sec.~\ref{sec:hkgmodel}, the transition to the high-field phase takes place at $\Bcrit\approx 7.4$, $10.1$ and $30.4$~T for $\ha$, $\hb$ and $\hc$, respectively (without quantum corrections taken into account). We also note that, for these field directions, the magnetization is saturated in the classical limit, even though this does not hold for finite $S$. This high-field phase displays gapped magnons, and their fate upon adding disorder is the main subject of this paper.

For $0<B<\Bcrit$ the system may display a complex evolution of phases, possibly with multiple different ordered states below $\Bcrit$.
For the specific parameter set of interest, the magnetization process has been studied in Ref.~\onlinecite{janssen2017}. For fields along one of the high-symmetry directions, there is a single finite-field transition at zero temperature between the canted zigzag and polarized phases. This transition is continuous when $\hb$ but of first order when $\ha$ or $\hc$.

\subsection{Bond disorder in Kitaev materials}

The mechanism for the emergence of Kitaev-dominated exchange models proposed by Jackeli and Khalliulin \cite{jackeli2009} is based on a combination of strong spin-orbit coupling and a particular exchange geometry, the latter leading to partial cancellations between multiple exchange paths which cause the Heisenberg interaction to be small. These cancellations as well as the processes leading to anisotropic interactions sensitively depend on the angles between different bonds (Ir-O-Ir or Ru-Cl-Ru). This is consistent with numerous ab-initio investigations which show that both pressure and strain lead to large changes of all microscopic couplings, including sign changes \cite{yadav2018,xu2020,kaib2021,wolf2022}.

Therefore, we can expect that crystallographic defects, locally changing the bond geometry, produce large inhomogeneities in the exchange couplings, i.e. large bond disorder. This can manifest in strong variations of the dominant couplings $K$ and $\Gamma$ as well as in the originally small Heisenberg coupling $J$ becoming large. We restrict our study to nearest neighbors and parametrize a single impurity on a bond of type $\gamma$ by
\begin{equation}
H^\mathrm{imp}_{\langle ij\rangle_\gamma} =
\vec{S}_i^{\,\mathrm{T}} \delta \mathcal{J}_\gamma\, \vec{S}_j,
\end{equation}
with
\begin{equation}
\delta\mathcal{J}_x=\begin{pmatrix}
\dJ{x}+\dK{x} & 0 & 0 \\
0 & \dJ{x} & \dG{x} \\
0 & \dG{x} & \dJ{x}
\end{pmatrix}
\end{equation}
for $\gamma=x$, and permutations.

Since reliable ab-initio studies of defect physics in Kitaev materials are lacking, we need to make assumptions on $H_{\mathrm{imp}}$. For isolated impurities, we will separately consider variations in $J$, $K$, and $\Gamma$, i.e we will discuss impurity types of $\dJ{x}$, $\dK{x}$, etc.
The results will guide us in selecting disorder distributions which induce low-energy states in the high-field phase, as suggested by experiments. Finally, by comparing our findings with available experimental data, we will propose plausible impurity models for \rucl\ and provide specific predictions.


\section{Spin-wave theory}
\label{sec:swt}

We will study disorder effects in the \hkg\ model in the semiclassical limit of large $S$, using the tool of spin-wave theory. This approach has been proven to yield semi-quantitative agreement with both experiments and numerics for $S=1/2$ in the ordered phases of Eq.~\eqref{Hamiltonian}. In this section we summarize the technical aspects involved; for details we refer the reader to Ref.~\onlinecite{wessel05}.
Quenched disorder will be treated exactly, by numerical computations for different disorder realizations on finite-size systems of $N=2L^2$ sites, with $L$ the linear system size, followed by proper disorder averaging for individual observables.

\subsection{Real-space spin-wave theory}

A spin-wave calculation is performed by expanding about a classical reference state, i.e., a spin product state which minimizes the classical energy. In the presence of magnetic field and disorder, we expect non-collinear states without translation invariance. Hence, the computation involves two steps, namely (i) finding the classical reference state by iterative minimization, and (ii) computing its excitation spectrum via spin-wave theory.

To perform the spin-wave expansion, we define a local coordinate system for each spin such that it is aligned to the $z$ axis, while the $x$ and $y$ axes can be arbitrary. Formally
\begin{equation}
\vec{S}_i^0=R_i(\theta_i,\phi_i) \,\vec{S}_i \, ,
\end{equation}
where the index $0$ denotes spin in the global laboratory frame and $R_i$ is a rotation matrix, which can be defined in different ways. Here we rotate the spin by $\theta_i$ about an axis $(-\sin\phi_i, \cos\phi_i,0)^\mathrm{T}$, which is perpendicular to both $\vec{S}^0$ and $\hat{z}$ and $\theta_i$ and $\phi_i$ are the spherical coordinates of the spin in the laboratory frame. Then the Hamiltonian can be rewritten as
\begin{equation}\label{Hamiltonian2}
H=
\sum_{\langle ij\rangle_\gamma}
\vec{S}_i^{\,\mathrm{T}} \tilde{\mathcal{J}}^\gamma_{ij}\, \vec{S}_j
+\sum_{\lllangle ij\rrrangle} \vec{S}_{i}^{\,\mathrm{T}} \tilde{J}_{3,ij} \vec{S}_{j}
-\sum_{i} \tilde{h}_i^{\,\mathrm{T}} \vec{S}_{i},
\end{equation}
with
$\tilde{\mathcal{J}}^\gamma_{ij}= R_i^{\,\mathrm{T}} \mathcal{J}^\gamma_{ij} R_j $,
$\tilde{J}_{3,ij}=J_3 \, R_i^{\,\mathrm{T}} R_j$
and
$\tilde{h}_i^{\,\mathrm{T}}= \vec{h}^{\,\mathrm{T}} R_i$.

In the local frame, each spin can be expressed in terms of Holstein-Primakoff bosons
\begin{equation}
\begin{split}
S_i^z&=S-a_i^\dagger a_i,\\
S_i^+&=\sqrt{2S-a_i^\dagger a_i}\,a_i, \\
S_i^-&=a_i^\dagger\sqrt{2S-a_i^\dagger a_i},
\end{split}
\end{equation}
and expanded in powers of $1/S$.
Inserting into the Hamiltonian the expansion can be organized as
\begin{equation}
H=H_0 + H_1 + H_2 + \ldots ,
\end{equation}
where indices refer to the number of bosonic operators in each term prior to any normal ordering operations. In the case of Hamiltonians that are bilinear in spin operators, $H_n$ carries a factor of $S^{2-n/2}$, but this changes in the presence of a magnetic field or higher-order spin exchange interactions.

$H_0$ represents the classical ground-state energy, and $H_1$ vanishes when the spin-wave expansion is performed with respect to a local extremum of the classical energy.
$H_2$ can be written as
\begin{equation}
H_2 = \vec{a}^\dag M \vec{a}
\end{equation}
up to constants, with a vector of operators $\vec{a}^\dag=(a^\dag_1,...,a^\dag_N,a_1,...,a_N)$ and $M$ a $2N \times 2N$ matrix of the form
\begin{equation}
M=\begin{pmatrix}
A & B \\
B^* & A^\mathrm{T}
\end{pmatrix},
\end{equation}
where $N$ is the total number of spins.

The spin-wave spectrum is obtained by diagonalizing $M$ using a bosonic Bogoliubov transformation, $\vec{a}=T\vec{b}$. Our numerical implementation follows the procedure described in Ref.~\onlinecite{wessel05}. Here we outline the key points of it. To begin with, the $\vec{b}$ operators in the diagonal basis must obey canonical bosonic commutation relations, which imposes the constraint
\begin{equation}
T^\dagger \Sigma T=\Sigma,
\qquad \text{with} \quad
\Sigma=\begin{pmatrix}
\mathbb{1}_N & 0 \\
0 & -\mathbb{1}_N
\end{pmatrix},
\end{equation}
and as a result $T$ is of the form
\begin{equation}\label{UVdef}
T=\begin{pmatrix}
U & V \\
V^* & U^*
\end{pmatrix},
\end{equation}
with $U$ and $V$ being $N \times N$ matrices. Then, from
$T^\dagger M T=\Omega$, where $\Omega$ is a diagonal matrix containing the eigenvalues of the Hamiltonian, it follows that
\begin{equation}
\Sigma M T=T \Sigma \Omega.
\end{equation}
Therefore the problem of diagonalizing the Hamiltonian is reduced to the diagonalization of the non-Hermitian matrix $\Sigma M$. Its eigenvalues are $\Sigma \Omega$, while $T$ is the matrix of its right eigenvectors.
In the case of degenerate eigenvalues, which is often the case when the Hamiltonian is expressed in real space, the eigenvectors corresponding to the same eigenvalue are not necessarily orthogonal and need to be explicitly orthogonalized.

As our algorithm doubles the degrees of freedom of the system, $\Sigma M$ has $2N$ eigenvalues that come in pairs of $\pm \tilde{\w}_i$. The physical excitation energies then are $\w_i=2|\tilde{\w}_i|$, and are $N$ in number, on par with the original degrees of freedom.


\subsection{Observables}

We use the saddle-point plus spin-wave calculation to extract a number of observables characterizing the magnetic state and its modifications from impurities. For most observables, we restrict our attention to the respective leading-order in $1/S$. Below, $S^0$ denote spin operators in the global laboratory frame.

The uniform magnetization is simply $\vec{m}=\sum_i \vec{S}^0_i$, which we calculate without quantum corrections. It is useful to quantify deviations of the moments from the field direction,
\begin{equation}\label{order_parameter}
m_\perp^\text{loc}=\sum_i |\vec{S}^0_{i,\perp} |,
\end{equation}
where $\vec{S}^0_{i,\perp}$ is the projection of each spin into the plane perpendicular to the field. The static spin structure factor
\begin{equation}
\mathcal{S}^{\alpha\beta}(\vec{q}) = \frac{1}{N}
\sum_{i,j}e^{-i\vec{q}\cdot(\vec{r}_i-\vec{r}_j)}
\langle S^{0,\alpha}_i S^{0,\beta}_j \rangle,
\end{equation}
where $\alpha, \beta=\{x_0,y_0,z_0\}$ are coordinates in the laboratory frame, signals ordering tendencies. As above, we restrict ourselves to the leading classical contribution and, in particular, we calculate the trace $\mathcal{S}(\vec{q})=\sum_\alpha S^{\alpha \alpha}(\vec{q})$.

Spin excitations can be deduced from the dynamic spin structure factor, accessible in inelastic neutron scattering experiments,
\begin{equation}
\mathcal{S}^{\alpha\beta}(\vec{q},\omega) = \frac{1}{N}
\sum_{i,j}e^{-i\vec{q}\cdot(\vec{r}_i-\vec{r}_j)}
\mathcal{S}^{\alpha\beta}_{ij}(\omega),
\end{equation}
with
\begin{equation}
\mathcal{S}^{\alpha\beta}_{ij}(\omega) =
\int \mathrm{d}t \, e^{i \omega t}
\langle S_i^{0,\alpha}(t) S_j^{0,\beta}(0)\rangle.
\end{equation}
Using the zero-temperature version of the fluctuation--dissipation theorem, $\mathcal{S}(\omega)= 2 \,\theta(\omega) \mathfrak{Im}\chi(\omega)$, and expressing the susceptibility $\chi(\omega)$ in a Lehmann representation, spin-wave theory yields in single-mode approximation
\begin{equation}
\mathcal{S}^{\alpha\beta}_{ij}(\omega) = S\pi
\sum_{k=1}^N
\big( u^\alpha_{ik} -i v^\alpha_{ik} \big)^*
\big( u^\beta_{kj} -i v^\beta_{kj} \big)
\delta(\omega-\omega_k),
\end{equation}
where $u^\alpha_{ik}=R^{\alpha x}_i (U_{ik}+V^*_{ik})$ and $v^\alpha_{ik}=R^{\alpha y}_i (U_{ik}-V^*_{ik})$, $R_i$ is the rotation matrix of $\vec{S}_i$, and $U$ and $V$ the matrices of Bogoliubov coefficients defined in Eq.~\eqref{UVdef}.
Unless otherwise noted, we will display results for the spin-unpolarized structure factor, $\mathcal{S}(\vec{q},\w) = \sum_\alpha \mathcal{S}^{\alpha\alpha}(\vec{q},\w)$, and we will also consider the momentum-integrated (i.e., local) structure factor
\begin{equation}
\mathcal{S}(\omega) = \frac{1}{N} \sum_{i} \mathcal{S}_{ii}(\omega)
\end{equation}
To display the numerical results we employ Lorentzian broadening with linewidth $\eta$,
\begin{equation}
\delta(\omega-\omega_k) \,\rightarrow\,
\frac{1}{\pi} \frac{\eta}{(\omega-\omega_k)^2+\eta^2}.
\end{equation}
Because of the Bogoliubov transformation, modes with $\omega_k\to 0$ have diverging spectral weight. Therefore we exclude all modes with $\omega_k<10^{-3} J$ from the plots.


\subsection{T-matrix theory for a single impurity}
\label{sec:tmatrix}

In the high-field phase, the changes to the spin-wave spectrum caused by a single impurity can be efficiently computed by scattering theory. Even though the T-matrix method has been applied to inhomogeneous systems \cite{brenig2013}, we will restrict ourselves to cases where the classical reference remains translation-invariant, i.e. does not develop a spin texture. The conditions for the latter will be discussed in Sec.~\ref{sec:1imp_symmetries} below.

In this T-matrix approach, spin waves scatter elastically off a local potential $V$ generated by the impurity. The real-space spin-wave propagator obeys
\begin{equation}
G_{ij}(\w) =G^0_{ij}(\w) + G^0_{i0}(\w) T(\w) G^0_{0j}(\w),
\end{equation}
with $G^0$ the bulk propagator, $i,j$ site indices, and the $T$ matrix given by
\begin{equation}
T(\w)=V(1-G^0_{0,0}(\w) \,V)^{-1},
\end{equation}
where $G^0_{0,0}$ is the local bulk propagator, with $r=0$ the impurity site. Importantly, in the models of interest, the impurity is spatially local in the basis of the original Holstein-Primakoff bosons.

The formation of bound states in the spectrum is signaled by poles in $T(\w)$. For a two-site unit cell and a bond defect inside the unit cell, the scattering potential $V$ necessarily has a $2\times 2$ matrix structure. Additionally, the Kitaev terms in Eq.~\eqref{Hamiltonian} generate anomalous terms in the spin-wave expansion, even in the fully polarized high-field phase, resulting in a $4\times 4$ matrix structure for the above equation. Poles of the $T$ matrix are then given by the zeroes of  $\det{\left(1-(\hat{G}^0\,\hat{V})_{mn}\right)}$, where $m$ and $n$ and the indices in sublattice space of the two neighboring spins connected by the defect bond.

The retarded bulk Green's function can be expressed in Nambu notation as
\begin{equation}
\hat{G}^{0}(\tilde{\omega})\!=\!
-i \!\int \!\mathrm{d}t e^{i\tilde{\omega} t} \, \theta(t)
\langle [ \bar{a}(t), \bar{a}^\dagger(t=0) ]\rangle,
\end{equation}
with $\bar{a}^\dag=(a^\dag_m,a^\dag_n,a_m,a_n)$.
It contains forward, backward and anomalous bosonic propagators and can thus take the block matrix form
\begin{equation}
\hat{G}^{0}(\tilde{\omega})=
\begin{pmatrix}
G^R(\tilde{\omega}) & F(\tilde{\omega})\\
\tilde{F}(\tilde{\omega}) & G^A(-\tilde{\omega}).
\end{pmatrix}
\end{equation}

In the particular case of a Heisenberg impurity, i.e. $\dJ{}\neq 0$ and $\dK{}=\dG{}=0$, by rewriting the impurity Hamiltonian in spin-wave theory terms as
$H_\mathrm{imp}=\bar{a}^\dagger \hat{V} \bar{a}$
we can extract the $4\times 4$ scattering matrix
\begin{equation}
\hat{V}=
\begin{pmatrix}
V & 0\\
0 & V\\
\end{pmatrix},
\quad
\text{with}
\quad
V=\dJ{}\,S
\begin{pmatrix}
-1 & 1 \\
1 & -1
\end{pmatrix}.
\end{equation}
Then, bound state energies can be found by solving the real part of the equation
\begin{equation}\label{T-matrixPole}
\begin{split}
\frac{1}{(\dJ{}\,S)^2}
+\frac{1}{\dJ{}\,S} \frac{1}{2} \Big( \mathcal{G}^R(\tilde{\omega})+\mathcal{G}^A(-\tilde{\omega}) \Big)&\\
+\frac{1}{4} \Big( \mathcal{G}^R(\tilde{\omega}) \mathcal{G}^A(-\tilde{\omega})
- \mathcal{F}(\tilde{\omega}) \tilde{\mathcal{F}}(-\tilde{\omega})\Big)&
=0,
\end{split}
\end{equation}
with
\begin{equation}
\mathcal{G}^R(\tilde{\omega})=G^{R}_{mm}(\tilde{\omega})-G^{R}_{mn}(\tilde{\omega})-G^{R}_{nm}(\tilde{\omega})+G^{R}_{nn}(\tilde{\omega})
\end{equation}
and equivalent expressions for $\mathcal{G}^A$, $\mathcal{F}$, $\tilde{\mathcal{F}}$.
The propagators can be found either by matrix inversion of $(\hat{G}^0)^{-1}$ or by diagonalization via a Bogoliubov transformation.
Since the above expressions are in Nambu notation, which double the degrees of freedom of the initial problem, the physical bound state energies will be $\omega=2|\tilde{\omega}|$, where $\tilde{\omega}$ solves Eq.~\eqref{T-matrixPole} for a given impurity strength.


\section{Isolated impurities}
\label{sec:1imp}

In this section, we study the physics of a single bond defect in an otherwise defect-free system. This corresponds to the limit of vanishingly small impurity concentration in a thermodynamically large system.

As will become clear below, there are cases of impurity-induced spontaneous symmetry breaking. Given that symmetry breaking requires the thermodynamic limit, a single impurity cannot, strictly speaking, spontaneously break symmetries. As our calculations are mean-field-based, they escape this rule. Hence, the corresponding statements below concerning symmetry breaking are valid for any small but finite impurity concentration.

Numerical computations are performed for a system of linear size $L=18$ unless noted otherwise.


\renewcommand{\arraystretch}{1.5}
\begin{table}[t]
\begin{center}
\begin{tabular*}{\columnwidth}{@{\extracolsep{\fill}}| l | c | c | c | c |}
\hline
 & $\ha$ & $\hb$ & $\hc$ & generic $\vec{B}$\\
\hline
Clean & \translation,\inversion{},\reflection{z} & \translation,\inversion{},\flip{z}  & \translation,\inversion{},\rotation,\reflection{x},\reflection{y},\reflection{z}   & \translation,\inversion{},\\
\hline\hline
\multicolumn{5}{|l|}{Single impurities} \\
\hline
$\dJ{z}$ &  \inversion{0},\reflection{z} & \inversion{0},\flip{z} & \inversion{0},\rotation,\reflection{x},\reflection{y}, \reflection{z}  &  \inversion{0} \\
\hline
$\dJ{x}$ &  \multirow{2}{*}{\centering{ \inversion{0},\reflection{z}}} & \multirow{2}{*}{\centering{ \inversion{0},\flip{z}}} &  \multirow{2}{*}{\inversion{0},\rotation,\reflection{x},\reflection{y}, \reflection{z}}   & \multirow{2}{*}{\centering{ \inversion{0}}} \\
$\dJ{y}$ &  & & & \\
 \hline
$\dK{z}$/$\dG{z}$ &  \inversion{0},\reflection{z} &  \inversion{0},\flip{z} &  \inversion{0},\reflection{z} & \inversion{0} \\
 \hline
$\dK{x}$/$\dG{x}$ & \multirow{2}{*}{\centering{ \inversion{0}}} &   \multirow{2}{*}{\centering{ \inversion{0}}}& \inversion{0},\reflection{x} & \multirow{2}{*}{\centering{ \inversion{0}}} \\
$\dK{y}$/$\dG{y}$ & & & \inversion{0},\reflection{z}  & \\
\hline\hline
\multicolumn{5}{|l|}{Impurity distributions} \\
\hline
$\dJ{}$ & \reflection{z} & \flip{z} &\rotation,\reflection{x},\reflection{y},\reflection{z} &  0 \\
\hline
$\dK{}$/$\dG{}$ &  0 &  0 &  0 & 0 \\
\hline
\end{tabular*}
\end{center}
\caption{Symmetries of the \hkg\, Hamiltonian with  a magnetic field applied in different directions  and no impurities, a single impurity of various types and random impurity distributions. Further explanation for each symmetry is provided in Sec.~\ref{sec:model:symmetries}.}
\label{table:symmetries}
\end{table}

\renewcommand{\dbltopfraction}{0.6}

\begin{figure}[t]
\includegraphics{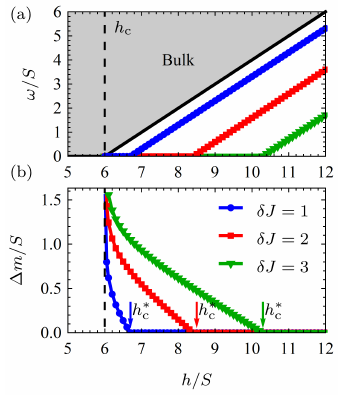}
\caption{
Heisenberg antiferromagnet with a single bond defect; data is shown for fields $h\geq\hcrit = 6S$ and three different impurity strengths.
(a) Energy of the magnon bound state. The bulk continuum is shaded; the vertical dashed line shows the bulk critical field $\hcrit$.
(b) Impurity-induced uniform magnetization $\Delta m=|\vec{m}_\text{clean}|-|\vec{m}_\text{dis}|$. Arrows indicate the critical field $\hcrit^\ast$ below which the impurity induces a texture with local canted order.}
\label{fig:Heisenberg}
\end{figure}

\begin{figure}[t]
\includegraphics[scale=1]{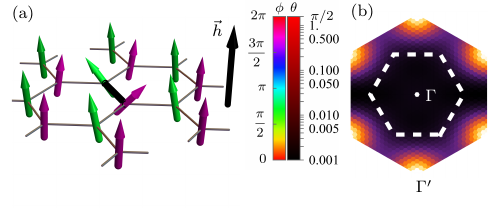}
\caption{
(a) Spin texture surrounding a bond impurity with $\dJ{}=2$ in the Heisenberg antiferromagnet at $h/S=7$.
The color of each spin encodes the angle $\phi$ of the spin projected on the plane perpendicular to the field, while the brightness shows the deviation $\theta$ of a spin from the field direction in logarithmic scale.
(b) Static spin structure factor, with the large peaks at momenta $\Gamma$ and $\Gamma'$ (the centers of the first and second Brillouin zones, respectively) corresponding to uniform bulk magnetization removed. The dashed line shows the boundary of the first Brillouin zone. The impurity induces a diffuse signal around $\Gamma'$.
}
\label{Heisenberg:texture}
\end{figure}


\begin{figure*}[t]
\includegraphics{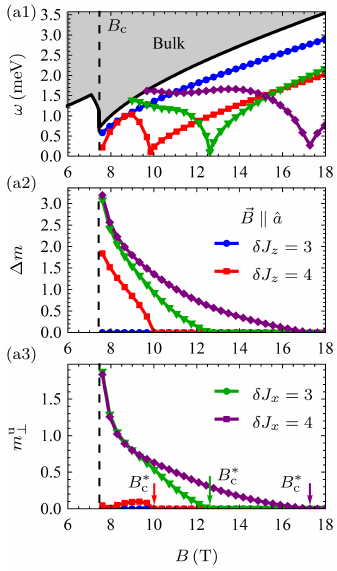}%
\includegraphics{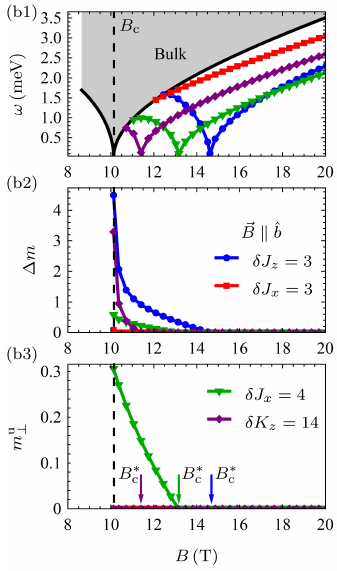}%
\includegraphics{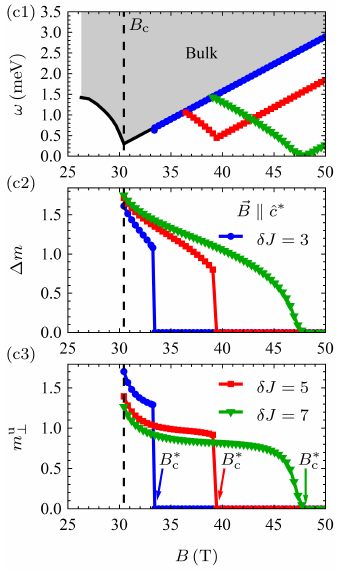}
\caption{
Characterization of type-II impurities, which induce textures via symmetry breaking only, in the {\hkg} model as a function of the magnetic field. Data is shown for three high-symmetry directions (two in plane, one out of plane) above the respectively bulk critical field. Impurities are chosen to have different strength (in units of meV) and, in the cases of $\ha$ and $\hb$, to be placed on two inequivalent bonds.
(a1-c1) Impurity-induced uniform magnetization $\Delta m=|\vec{m}_\text{clean}|-|\vec{m}_\text{dis}|$;
(a2-c2) Magnetization transverse to the applied field $m_\perp^u=|\vec{m}_{\perp\vec{h}}|$.
(a3-c3) Energy of the magnon bound state.
}
\label{fig:broken}
\end{figure*}

\begin{figure}[t]
\centering
\includegraphics{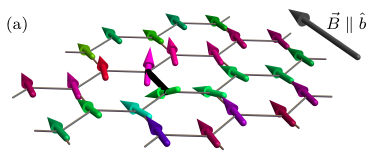}
\includegraphics{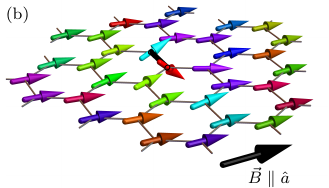}
\caption{
Symmetry-breaking textures for
(a) $\vec{B}=15.55\,\mathrm{T}\,\hat{b}$ and $\dJ{z}=4$~meV, which breaks \flip{z}, and
(b) $\vec{B}=9.67\,\mathrm{T}\,\hat{a}$ and $\dJ{z}=4$~meV, which breaks \reflection{z}.
For the spin color scheme see Fig.~\ref{Heisenberg:texture}.
}
\label{fig:texture_broken}
\end{figure}

\begin{figure}[t]
\includegraphics{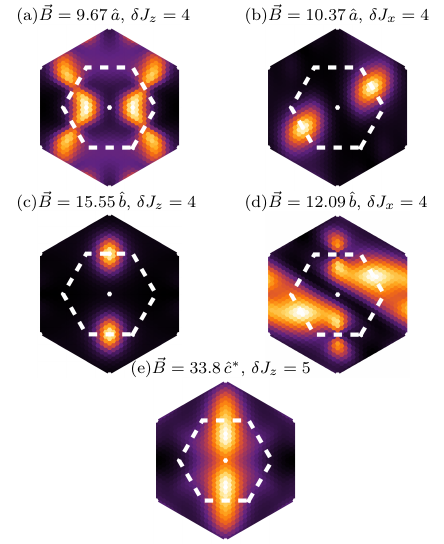}
\caption{
Static structure factor as in Fig.~\ref{Heisenberg:texture}(b), but here for the {\hkg} model with different symmetry-breaking textures with magnetic fields in one of the three high-symmetry directions and, for in-plane fields, two inequivalent defect bonds. The type of magnetic order of each texture depends on the combination of field direction and placement of the impurity, for details see text.
Magnetic fields are quoted in T and impurity strengths in meV.
}
\label{fig:SSFbroken}
\end{figure}

\renewcommand{\topfraction}{0.7}
\begin{figure*}
\includegraphics{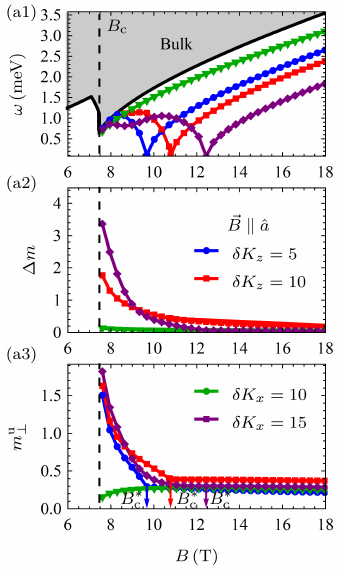}%
\includegraphics{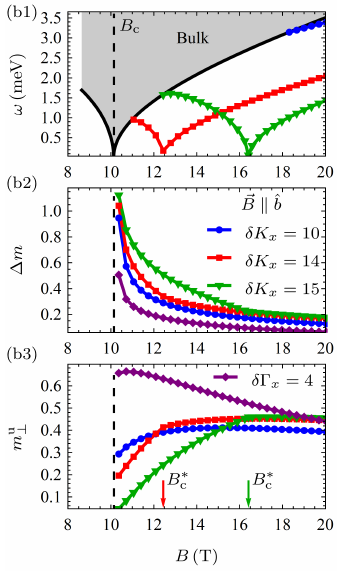}%
\includegraphics{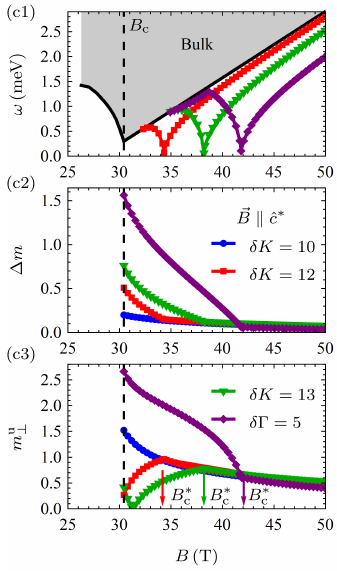}
\caption{
As in Fig.~\ref{fig:broken}, but now for type-I impurities in the {\hkg} model inducing symmetric textures. Remarkably, many cases lead to symmetry-breaking textures at low fields in addition to the higher-field symmetric textures.
}
\label{fig:unbroken}
\end{figure*}

\begin{figure}[tb]
\centering
\includegraphics{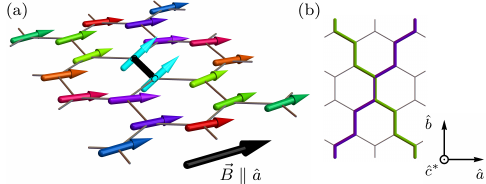}
\caption{(a) Texture with \reflection{z} symmetry hosting two zigzag modulations, as sketched in (b). This corresponds to double-Q order, leading to two pairs of peaks in the structure factor shown in Fig~\ref{fig:SSFunbroken}(a) below. Data was generated for field $\vec{B}=13.82\,\mathrm{T}\,\hat{a}$ and impurity $\dK{z}=10$~meV.
Color scheme for spins as in Fig.~\ref{Heisenberg:texture}.}
\label{fig:textures_unbroken}
\end{figure}

\begin{figure}[tb]
\includegraphics{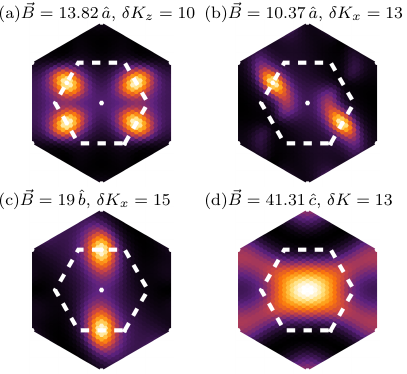}
\caption{
Static structure factors as in Fig.~\ref{Heisenberg:texture}(b), but here for symmetric textures for all combinations of field direction and impurity orientation. (a,d) are symmetric under \reflection{z} and \inversion{0}, and
(b,c) only under \inversion{0}.
}
\label{fig:SSFunbroken}
\end{figure}

\renewcommand{\topfraction}{0.6}
\begin{figure*}[tb]
\includegraphics{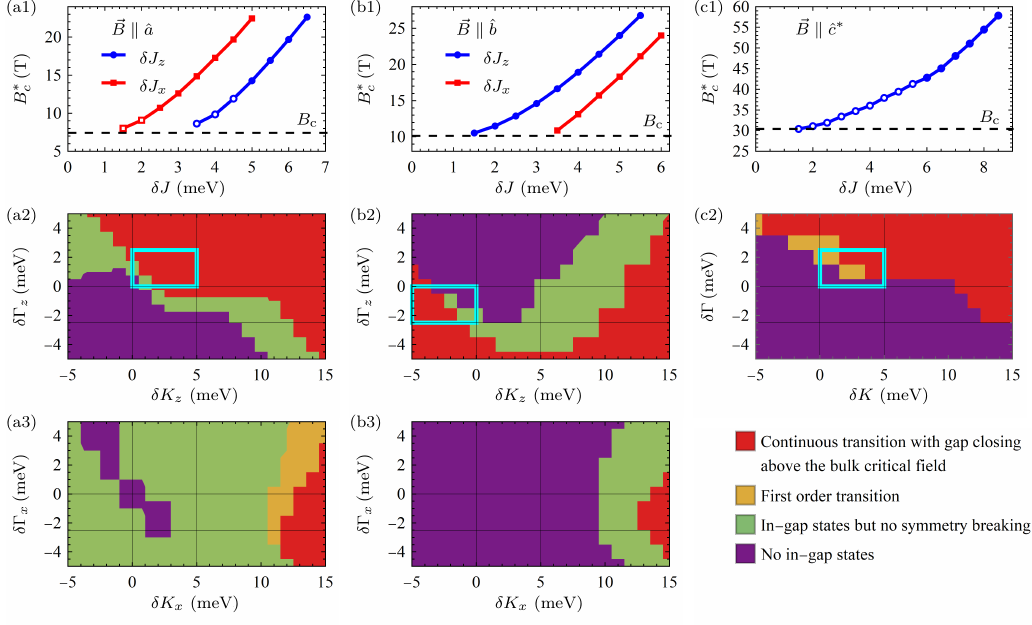}
\caption{
Relation between impurity strengths and low-energy in-gap states for the {\hkg} model.
(a1-c1) Transition field $h_c^*$ for texture formation as function of the strength of Heisenberg impurities. Filled (open) symbols denote  continuous (first-order) transitions.
(a2-b3) Existence of symmetry breaking and in-gap states for combined $\dK{}$ and $\dG{}$ impurities on a single bond, see legend for color codes. Regions of interest for Sec.~\ref{sec:manyimp} are highlighted in cyan, for details see text.
}
\label{fig:KG_comb}
\end{figure*}

\begin{figure}[tb]
\includegraphics[scale=1]{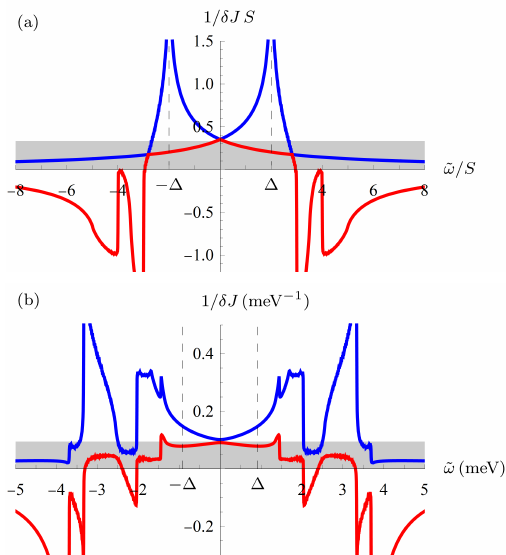}
\caption{
Solutions to the real part of Eq.~\eqref{T-matrixPole} for poles of the T-matrix for a single Heisenberg impurity. Each line corresponds to one of the two solutions of the quadratic equation, and bound-state energies are $\w=2|\tilde{\w}|$ for any $\dJ{}$.
(a) In the Heisenberg model (shown for $h/S=10$) the divergence at the band edge $\w=\pm\Delta$ leads to bound states in the gap for arbitrarily weak impurities.
(b) In the {\hkg} model (shown for $\vec{B}=13.82\,\mathrm{T}\,\hat{a}$) the Bogoliubov coefficients cancel the divergence at $\Delta$, requiring a minimum impurity strength to induce a bound state.
Energies are in units of $J$ in (a) and meV in (b).
The shaded regions correspond to unphysical solutions, see text.
}
\label{fig:T-matrix}
\end{figure}


\subsection{Bond defects and textures}
\label{sec:1imp_symmetries}

A defect will generically change the magnetic order in its vicinity, and those changes can be separated into changes of (i) moment directions, and (ii) moment amplitudes. For classical spins, moment amplitudes are fixed, and therefore we focus here on changes of moment directions, which correspond to impurity-induced spin textures. In the following we discuss the high-field phase where the clean-limit classical ground state is uniform and, for a field $\hpc$ or $\hc$, polarized in the field direction; many of the considerations apply to other bulk phases as well.

As detailed below, two cases can be distinguished and correspond to two classes of impurity problems:
Type-I impurities generically create a texture independent of their impurity strength and for any $B>\Bcrit$. Such textures result from the fact that the presence of the impurity locally breaks the symmetries protecting the uniform state at the Hamiltonian level. In contrast, type-II impurities do not generically create a texture, but they may induce a texture via the spontaneous breaking of symmetries. The latter typically happens for impurity strengths beyond a critical value. Consequently, we will distinguish two types of textures as well, namely symmetric and symmetry-breaking textures. We note that even type-I impurities can generate symmetry-breaking textures, i.e., tuning the impurity strength can induce a transition from a symmetric to a (less symmetric) symmetry-breaking texture.

Importantly, the symmetries protecting the uniform state can be different: For the Heisenberg model, it is the U(1) spin rotation about the field axis, even in the absence of translation symmetry. For \hkg, it is translation symmetry provided that the field is $\hpc$ or $\hc$; if translation is broken, the uniform state is still preserved if the field is parallel or perpendicular to the Kitaev axis of the impurity which applies, e.g., to $\dK{z}$ or $\dG{z}$ with $\hb$. The case of $\dJ{}$ impurities is special: Because of its local spin rotation symmetry, it leaves the uniform state intact for any of the field directions $\hpc$ or $\hc$.

On a technical level, type-I (type-II) impurities are those which do (do not) generate terms linear in bosonic operators, respectively, when performing a spin-wave expansion about the clean-limit classical state. Such linear terms signal Bose condensation which is equivalent to a change of the reference state. In practice, these cases require a full computation of the (now inhomogeneous) reference state by classical minimization.
In the high-field phase, linear terms do not occur for Heisenberg impurities, but generically occur for Kitaev and $\Gamma$ impurities unless the rotation matrices obey
\begin{equation}
\label{condition}
R^{\gamma\alpha}=R^{\gamma\beta}=0
\quad\text{or}\quad
R^{\gamma\gamma}=0
\end{equation}
for $\alpha,\beta\neq\gamma$ with $\gamma$ corresponding to the impurity bond. These conditions are consistent with the symmetry statements above.

Both types of textures will be discussed in detail below. In the context of symmetry-breaking textures, it is important to note that the symmetries available to break are strongly reduced compared to the clean bulk, Sec.~\ref{sec:model:symmetries}, since many symmetries are explicitly broken by adding an impurity. Obviously, translation symmetry is broken and, as a result, spatial inversion remains a symmetry only when performed about the center of the impurity bond, which we label \inversion{0}. Symmetries acting on both the lattice and the spins depend on the direction of the magnetic field and the type of the impurity. The relevant symmetries are summarized in Table \ref{table:symmetries}.

\subsection{Heisenberg limit}
\label{sec:1imp_Heisenberg}

As a warmup, we discuss the Heisenberg antiferromagnet, i.e. a model with $K=\Gamma=\Gamma'=0$, with bond disorder in its high-field phase. The absence of spin-orbit coupling reduces the complexity of the problem compared to the full {\hkg} model \eqref{Hamiltonian}. The Heisenberg model displays SU(2) spin symmetry which is reduced to U(1) in the presence of a magnetic field. The transition between the fully polarized high-field state and the symmetry-broken canted antiferromagnet takes place at a field $\hcrit$ which is given by $2 J S z$ where $z$ is the number of nearest neighbors. In this subsection, we employ energy units of $J=1$ and a field $h$ in the same units.

The fully polarized state is an eigenstate of $H$, which remains true in the presence of a defect bond, since it preserves the U(1) symmetry, and, as a result, a bond defect represents a type-II impurity. The defect modifies the spin-wave spectrum, which can be calculated using the T-matrix formalism, or explicitly using the real-space spin-wave approach. A defect with strength $\dJ{}>0$ locally decreases the energy cost for spin-flip excitations, and consequently we find a magnon bound state below the bulk continuum, Fig.~\ref{fig:Heisenberg}(a). In fact, a bound state occurs for any $\dJ{}>0$, as is explained in Sec.~\ref{tmatrixana} below. The bound state's energy depends on both field and impurity strength: a stronger impurity leads to a bound state at lower energy. Conversely, an impurity with $\dJ{}<0$ increases the spin-flip energy cost and creates a magnon anti-bound state above the bulk continuum (not shown).

Upon lowering the magnetic field, the magnon bound state for $\dJ{}>0$ decreases its energy, Fig.~\ref{fig:Heisenberg}(a), until it eventually condenses at $\hcrit^\ast(\dJ{}) > \hcrit$. This implies an instability of the assumed polarized state. Indeed, by minimizing the classical energy for $\hcrit<h<\hcrit^\ast$, we find that the classical ground state includes a spin texture with broken U(1) and \inversion{0} symmetries.
We quantify the impact of the impurity to the classical state by the deviation of the uniform magnetization from full polarization, $\Delta m=| \vec{m}_\text{clean}|-|\vec{m}_\text{dis}|$. As seen in Fig.~\ref{fig:Heisenberg}(b), $\Delta m$ onsets at $\hcrit^*$ and increases upon approaching the bulk transition.
For $h\leq \hcrit^\ast$, the magnon bound state has zero energy, Fig.~\ref{fig:Heisenberg}(a), i.e., it is a Goldstone mode related to the spontaneous breaking of the U(1) symmetry by the texture.

The textured state can be further analyzed by studying the real-space spin configuration, as portrayed in Fig.~\ref{Heisenberg:texture}(a). The impurity locally induces a canted antiferromagnetic order with a canting angle $\theta$ that decays exponentially with distance from the impurity, see Appendix~\ref{app:decay} for details. The static structure factor of the textured state, Fig.~\ref{Heisenberg:texture}(b), shows broad intensity around the $\Gamma'$ points of the second Brillouin zone, where the Bragg peaks of the N\'eel antiferromagnet are located, consistent with the texture being a precursor to the low-field ordered state.


\subsection{Impurities with symmetry-breaking textures}
\label{sec:1imp:broken}

We now turn to the full {\hkg} model, where---compared to the Heisenberg model---spin-orbit coupling reduces the symmetries and renders the field directions inequivalent. We will employ the parameter set specified in Sec.~\ref{sec:hkgmodel}, and we will quote energies in meV and magnetic fields in T.
This subsection will focus on impurities of type II, which may lead to textures only via symmetry breaking. These are $\dJ{}$ for any direction and $\dK{z}$ or $\dG{z}$ for $\hb$. We will present results for magnetic fields in high-symmetry directions, for which the remaining discrete symmetries of the Hamiltonian can be read off from Table~\ref{table:symmetries} for each type of defect.

The behavior of these impurities upon field evolution can be seen in Fig.~\ref{fig:broken} and has some characteristics in common with the Heisenberg case discussed in Sec.~\ref{sec:1imp_Heisenberg}: The classical state remains polarized at high fields, while an antiferromagnetic impurity may induce in-gap states in the magnon spectrum  and lead to the formation of textures via symmetry breaking upon approaching the bulk critical field. Depending on parameters, the corresponding transition at $\Bcrit^\ast$ can be continuous, Fig.~\ref{fig:broken}(a,b) or of first order, Fig.~\ref{fig:broken}(c).

One significant effect of the reduced symmetry is that symmetry-breaking textures only break discrete symmetries and hence do not produce Goldstone modes for $\Bcrit<B<\Bcrit^\ast$. Instead, if the transition at $\Bcrit^\ast$ is continuous, then the bound state's energy only softens at $\Bcrit^*$. If the transition is of first order, however, there is no mode softening, Fig.~\ref{fig:broken}(c1). Another difference to the Heisenberg case is that impurities may induce bound states which do \textit{not} condense upon lowering the field, but instead merge with the continuum, Fig.~\ref{fig:broken}(b1). The reasons behind this will be discussed in more detail in Sec.~\ref{sec:1imp_KG}.

When textures are formed, they do so by spontaneously breaking the discrete symmetries listed in Table~\ref{table:symmetries}. The result is a canting of spins that forms a zigzag pattern and decays exponentially with distance from the impurity bond, Fig.~\ref{fig:texture_broken}. For the high-symmetry field directions considered, where the clean classical state is polarized in field direction, the textures imply a deviation from full polarization, $\Delta m \neq 0$, Fig.~\ref{fig:broken}(a2-c2), as well as finite spin components transverse to the field in many cases, Fig.~\ref{fig:broken}(a3-c3).

In Fig.~\ref{fig:SSFbroken} we can see the static structure factors of textures for all inequivalent combinations of field and impurity directions. In Fig.~\ref{fig:SSFbroken}(b,c,e), the textures showcase a zigzag pattern with an orientation corresponding to that of the low-field clean case. In particular, when $\ha$ or $\hc$, the low-field clean ground state has two or three zigzag orientations respectively that are degenerate, and an impurity (i.e. its bond orientation) in the high-field regime uniquely chooses one of them, as is the case in Fig.~\ref{fig:SSFbroken}(b,e). In contrast, in the case of textures portrayed in Fig.~\ref{fig:SSFbroken}(a,d), the impurity does not fit in the zigzag orientations favored by the field and, therefore, a compromise is achieved by a texture with a more complicated structure, such as mixing of different zigzag directions. More on the real-space configurations of textures is discussed in Appendix \ref{app:textures}.


\subsection{Impurities with symmetric textures}
\label{sec:1imp:unbroken}

In this section, we discuss type-I impurities that create textures without symmetry breaking and regardless of impurity and field strength. Their field-dependent characteristics are displayed in Fig.~\ref{fig:unbroken}. Notably, the presence of textures does not guarantee spin-wave modes in the bulk gap for any impurity of this category. Instead, the existence of such in-gap states depends on the impurity sign and strength, as for type-II impurities. This is illustrated in Fig.~\ref{fig:unbroken}(a1-c1) for different combinations of impurity type and field direction; a texture without in-gap state occurs, e.g., for a $\dG{x}$ impurity and $\hb$. For cases where an in-gap state exists, it may disappear into the bulk for smaller fields, or it may condense at a field $\Bcrit^*>\Bcrit$, at which point the texture transits to one with broken symmetry, not unlike what happens for type-II impurities. The conditions on the presence of in-gap states and symmetry breaking are further analyzed in the next section.

Symmetric textures are again characterized by a finite impurity-induced magnetization, Fg.~\ref{fig:unbroken}(a2-c2), and magnetization components perpendicular to the applied field, Fig.~\ref{fig:unbroken}(a3-c3). Both of these quantities now appear regardless of the sign and strength of the impurity and extend over the entire field range, in contrast to symmetry-breaking textures. The transition from symmetric to symmetry-breaking textures upon lowering the field leads to non-analytic changes in these quantities.

Since there is no symmetry breaking involved in this mechanism for texture formation, textures preserve the symmetries listed in Table~\ref{table:symmetries}. An example is shown in Fig.~\ref{fig:textures_unbroken} being symmetric under \reflection{z} and \inversion{0}. Spin structure factors of states with symmetric textures are shown in Fig.~\ref{fig:SSFunbroken}. The zigzag patterns exhibited by different textures result from a combination of the preserved symmetries and the states allowed by each field direction. For instance, while the textures portrayed in Fig.~\ref{fig:SSFunbroken}(b,c) are both symmetric under \inversion{0}, their zigzag orientations differ, owing to the different field directions. On the other hand, Fig.~\ref{fig:SSFunbroken}(a) shows a state with double-Q order reflecting the presence of two zigzag modulations, Fig.~\ref{fig:textures_unbroken}. Such a state is allowed by \reflection{z} symmetry and, remarkably, does not exist in the clean case, but can be only induced by impurities.


\subsection{Impurity strengths and in-gap states}
\label{sec:1imp_KG}

The experimental data of Refs.~\onlinecite{hentrich2020,baek2017,baek2020} suggest the presence of impurity-induced magnetic in-gap states in {\rucl} for a field range above $\Bcrit$. We therefore use our single-impurity analysis to identify under which circumstances an impurity leads to a mode near $\w=0$; as we have seen such (near-)zero modes occur at continuous impurity-induced transitions toward symmetry-breaking textures. This identification will guide us in proposing experimentally relevant disorder distributions in Sec.~\ref{sec:manyimp} below.

\subsubsection{Numerical results}

In the SU(2)-symmetric Heisenberg case, our numerics in Sec.~\ref{sec:1imp_Heisenberg} revealed that all impurities with $\dJ{}>0$ induce in-gap states, which eventually condense at $\Bcrit^\ast$ forming a texture. Both the energetic distance of the in-gap mode from the bulk gap and $(\Bcrit^\ast-\Bcrit)$ increase monotonically with $\dJ{}$; this will be rationalized below.

In the {\hkg} model, however, the behavior is significantly more complicated, with an overview given in Fig.~\ref{fig:KG_comb} for the bulk model parameters specified in Sec.~\ref{sec:model}. Heisenberg impurities (Fig.~\ref{fig:KG_comb}(a1,b1,c1)) lead to a field-induced gap-closing transition only beyond a certain antiferromagnetic strength, which depends on the orientation of the impurity bond unless $\hc$. In contrast, weaker impurities produce an in-gap state which merges with the bulk continuum but does not soften upon approaching the bulk critical field, such as the $\dJ{x}=3$ case in Fig.~\ref{fig:broken}(b1). Also, there are instances where the transition is of first order with the in-gap state remaining at finite energy. These are represented by open symbols in Fig.~\ref{fig:KG_comb}(a1,c1).

For $\dK{}$ and $\dG{}$ impurities the effects of bulk and impurity anisotropies intertwine, and we choose to analyze isolated defects which combine $\dK{}$ and $\dG{}$ on the same bond, with results in Figs.~\ref{fig:KG_comb}(a2-b3). These panels indicate, depending on field direction and impurity orientation, for which impurity parameters in-gap states occur (in the field range $\Bcrit<B<2\Bcrit$), and whether a transition to a symmetry-breaking texture occurs upon lowering the field, either continuous or of first order. The general trend is that sufficiently strong impurities which oppose a homogeneously magnetized state induce in-gap states; these are impurities with $\dK{}>0$ for any field direction, $\dG{}>0$ for $\hc$, $\dG{z}>0$ for $\ha$ and $\dG{z}<0$ for $\hb$.

Considering the results in Fig.~\ref{fig:KG_comb}, we see that impurity-induced mode softening above $\hcrit$ often requires impurities which are strong compared to their bulk values. However, we can identify some combinations of weak $\dK{}$ and $\dG{}$, marked with boxes in Fig.~\ref{fig:KG_comb}(a2-c2), which lead to mode softening, namely $\dK{}>0$ and $\dG{}>0$, i.e. weaker ferromagnetic $K$ and stronger $\Gamma$, and $\dK{}<0$ and $\dG{}<0$, i.e. stronger ferromagnetic $K$ and weaker $\Gamma$. In the first case, impurity-induced soft modes arise $\ha$ and $z$-bond impurities as well for $\hc$ regardless of the impurity orientation, while this happens for $\hb$ and $z$-bond impurities in the second case.

\subsubsection{Analytical T-matrix interpretation}
\label{tmatrixana}

In the absence of textures, impurity-induced in-gap states can be predicted via the T-matrix formalism of Sec.~\ref{sec:tmatrix}, which amounts to finding solutions to Eq.~\eqref{T-matrixPole}. Here, we will use this to interpret the circumstances under which bound states may exist and whether they condense.

We start from the Heisenberg antiferromagnet, again as a useful benchmark. In the polarized phase there exist no anomalous terms in the  spin-wave expansion and, as a result, Eq.~\eqref{T-matrixPole} contains only the normal propagators $\mathcal{G}^R(\tilde{\omega})$ and $\mathcal{G}^A(-\tilde{\omega})$. These can be found by a Fourier transformation and subsequent diagonalization of $2\times 2$ matrices, leading to
\begin{equation}
\mathcal{G}^{R/A}(\tilde{\omega})=\frac{1}{N} \sum_{l,\vec{k}}
|u_{l,\vec{k}}|^2 \frac{1}{\tilde{\omega}\pm i\eta-\tilde{\omega}_{l,{\vec{k}}}}
\end{equation}
where $u_{l,\vec{k}}=U_{1l}-e^{\Delta\vec{r}\cdot\vec{k}} U_{2l}$ with $U$ the unitary matrix that diagonalizes the Hamiltonian, 1,2 sublattice indices, $l$ the band index, $\vec{k}$ momentum, $\eta$ an infinitesimal imaginary part, and $\Delta\vec{r}$ the vector of the impurity bond.

In Fig.~\ref{fig:T-matrix}(a) we present the solution to Eq.~\eqref{T-matrixPole} for a fixed magnetic field and showcase the bound-state energy as a function of impurity strength $\dJ{}$. The propagators $\mathcal{G}^{R/A}(\tilde{\omega})$ have been calculated numerically from the equation above for an impurity on the $z$ bond.
We observe that the solution exhibits a divergence at the band edge. This can be qualitatively understood from a long-wavelength approximation to $\mathfrak{Re}\mathcal{G}^{R/A}(\tilde{\omega})$: Near the dispersion minimum at $\vec{k}=0$ we expand the dispersion as $\tilde{\omega}_k\approx \Delta+ b k^2$, with $b$ a velocity. The momentum summation is performed in the continuum limit, $\frac{1}{N}\sum_k\rightarrow \int \frac{\mathrm{d}^2 k}{(2\pi)^2}$, and in polar coordinates with a UV cut-off $\Lambda$. This yields
\begin{equation}
\mathfrak{Re}\mathcal{G}^{R/A}_\mathrm{H}(\tilde{\omega})\sim
\int_0^\Lambda \!\frac{\mathrm{d}k}{2\pi}\,
\frac{k}{\tilde{\omega}-\Delta-b k^2}.
\end{equation}
Power counting shows that the above integral diverges logarithmically at $\tilde{\omega}=\Delta$ in the case of $\mathfrak{Re}\mathcal{G}^{R}_\mathrm{H}(\tilde{\omega})$ and at $\tilde{\omega}=-\Delta$ for $\mathfrak{Re}\mathcal{G}^{A}_\mathrm{H}(-\tilde{\omega})$, in agreement with our exact solution to the T-matrix poles in Fig.~\ref{fig:T-matrix}.
The physical consequence of this divergence is that arbitrarily weak antiferromagnetic impurities induce a bound state in the gap, close to the band edge, as is also known for generic impurities in quadratically dispersing systems in two space dimensions.

Increasing the impurity strength, Fig.~\ref{fig:T-matrix}(a) predicts a decrease in the energy of the bound state, which eventually goes to zero for a $\dJ{\text{crit}}$, in agreement with our real-space calculation in Sec.~\ref{sec:1imp_Heisenberg}. This mode condensation at $\dJ{\text{crit}}$ implies the transition to a textured symmetry-broken state. Therefore, the present T-matrix calculation---which assumes a homogeneous reference state---cannot predict the behavior for $\dJ{}>\dJ{\text{crit}}$; the poles of the T matrix in this regime (shaded in Fig.~\ref{fig:T-matrix}) are unphysical.

For ferromagnetic impurities ($\dJ{}<0$), Fig.~\ref{fig:T-matrix}(a) predicts a anti-bound state above the bulk continuum, the energy of which increases as the impurity gets stronger. This is a valid solution, since $\dJ{}<0$ impurities never form a texture when the bulk is polarized.
Note that the additional solutions shown inside the continuum do not represent true poles of the T matrix (they indicate a
reshuffling of the bulk spectrum).

We now turn to the \hkg\ model where anomalous terms in the spin-wave Hamiltonian exist, even when expanded about the polarized state. As a result, Eq.~\eqref{T-matrixPole} for the poles of the T-matrix contains both normal and anomalous propagators. These can be calculated by successively performing Fourier and Bogoliubov transformations, resulting in
\begin{align}
&\mathcal{G}^R(\tilde{\omega})=\frac{1}{N}\!\sum_{l,\vec{k}}
\bigg[
\frac{|u_{l,\vec{k}}|^2}{\tilde{\omega}+i\eta - \tilde{\omega}_{l,\vec{k}}}
+\frac{|v_{l,\vec{k}}|^2}{-\tilde{\omega}-i\eta - \tilde{\omega}_{l,\vec{k}}}
\bigg],\\
&\mathcal{G}^A(-\tilde{\omega})=\!\frac{1}{N}\!\sum_{l,\vec{k}}\!
\bigg[
\frac{|v_{l,\vec{k}}|^2}{\tilde{\omega}+i\eta - \tilde{\omega}_{l,\vec{k}}}
\!+\!\frac{|u_{l,\vec{k}}|^2}{-\tilde{\omega}-i\eta - \tilde{\omega}_{l,\vec{k}}}
\bigg],\\
&\mathcal{F}(\tilde{\omega})=\!\frac{1}{N}\sum_{l,\vec{k}}
\bigg[
\frac{\tilde{u}_{l,\vec{k}} v_{l,\vec{k}}}{\tilde{\omega}+i\eta - \tilde{\omega}_{l,\vec{k}}}
+\frac{u_{l,\vec{k}} \tilde{v}_{l,\vec{k}}}{-\tilde{\omega}-i\eta - \tilde{\omega}_{l,\vec{k}}}
\bigg],\\
&\tilde{\mathcal{F}}(\tilde{\omega})=\!\frac{1}{N}\sum_{l,\vec{k}}
\bigg[
\frac{u^\ast_{l,\vec{k}} \tilde{v}^\ast_{l,\vec{k}}}{\tilde{\omega}+i\eta - \tilde{\omega}_{l,\vec{k}}}
+\frac{\tilde{u}^\ast_{l,\vec{k}} v^\ast_{l,\vec{k}}}{-\tilde{\omega}-i\eta - \tilde{\omega}_{l,\vec{k}}}
\bigg],
\end{align}
with
$u_{l,\vec{k}}=U_{1l,\vec{k}}-e^{-i \Delta\vec{r}\cdot\vec{k}} U_{2l,\vec{k}}$,
$\tilde{u}_{l,\vec{k}}=U_{1l,\vec{k}}-e^{i \Delta\vec{r}\cdot\vec{k}} U_{2l,\vec{k}}$ and similar expressions for $v_{l,\vec{k}}$, $\tilde{v}_{l,\vec{k}}$, where $U_{\vec{k}}$ and $V_{\vec{k}}$ are matrices of Bogoliubov coefficients as defined in Eq.~\eqref{UVdef}, but in a momentum-space basis.

The resulting solutions of Eq.~\eqref{T-matrixPole} are illustrated in Fig.~\ref{fig:T-matrix}(b), for $\vec{B}=14.82\,\mathrm{T}\,\hat{a}$ and a defect on a $z$ bond, with the two curves corresponding to the two solutions of the quadratic equation. Contrary to the Heisenberg case, there is no divergence at the band edge $\Delta$, meaning that a minimum impurity strength is required  to induce a bound state in the gap, in agreement with our real-space numerics.
On a technical level, the absence of this divergence is rooted in the Bogoliubov coefficients. The magnon spectrum of the \hkg\ model in the high-field phase has two minima at $\vec{k}=\pm \vec{k}_M$ and, therefore, to perform a low-wavelength approximation each integral is split in two, and the lowest band is approximated by $\tilde{\omega}_{l_1,\vec{k}}\approx \Delta+ b (\vec{k}\pm \vec{k}_M)^2$. The terms containing the Bogoliubov coefficients are also Taylor expanded as
$u_{l_1,\vec{k}}=u_0 + \vec{u}_1\cdot(\vec{k}\pm \vec{k}_M)$ and $v_{l_1,\vec{k}}=v_0 + \vec{v}_1\cdot(\vec{k}\pm \vec{k}_M)$ etc.
Then, the retarded propagator, after renaming
$\vec{k}\pm \vec{k}_M \rightarrow \vec{k}$,
can be written as
\begin{equation}
\begin{split}
\mathfrak{Re} \mathcal{G}^R_\mathrm{HK\Gamma}(\tilde{\omega})&=
\int \!\frac{\mathrm{d}^2k}{(2\pi)^2}\,
\frac{2\,|u_0+\vec{u}_1\cdot\vec{k}|^2}{\tilde{\omega}-\Delta-b k^2}\\
&-\int \!\frac{\mathrm{d}^2k}{(2\pi)^2}\,
\frac{2\,|v_0+\vec{v}_1\cdot\vec{k}|^2}{\tilde{\omega}+\Delta+b k^2}.
\end{split}
\end{equation}
While these integrals diverge at $\tilde{\omega}=\pm\Delta$ in the presence of non-zero coefficients $u_0$ and $v_0$, similar to the Heisenberg case, this divergence is cured if $u_0=v_0=0$, consistent with Fig.~\ref{fig:T-matrix}(b).
Similar arguments can be made for $\mathfrak{Re} \mathcal{G}^A_\mathrm{HK\Gamma}(-\tilde{\omega})$, but, when it comes to the anomalous propagators $\mathcal{F}(\tilde{\omega})$ and $\tilde{\mathcal{F}}(\tilde{\omega})$, divergences must cancel in both the real and imaginary parts.

The blue curve approaching $\tilde\omega=0$ for increasing impurity strength again implies bound-state condensation at $\dJ{\text{crit}}$.
As with the Heisenberg model, T-matrix poles obtained for larger positive $\dJ{}$ (shaded) are unphysical because the assumption of a homogeneous reference state is violated. Additionally, our real-space calculations show the formation of a texture via a first-order transition in certain cases (Fig.~\ref{fig:KG_comb}(a1,c1)). Such behavior cannot be captured by the T-matrix calculation, as the first-order transition preempts the bound-state condensation.

To conclude, the T-matrix results for the magnon bound states are consistent with those of the real-space calculations as long as the underlying magnetic state is uniform. Consequently, the T-matrix calculation can predict a continuous transition into a textured state via magnon condensation, but not a first-order transition.


\renewcommand{\topfraction}{0.7}

\begin{figure}[t]
\includegraphics[scale=1]{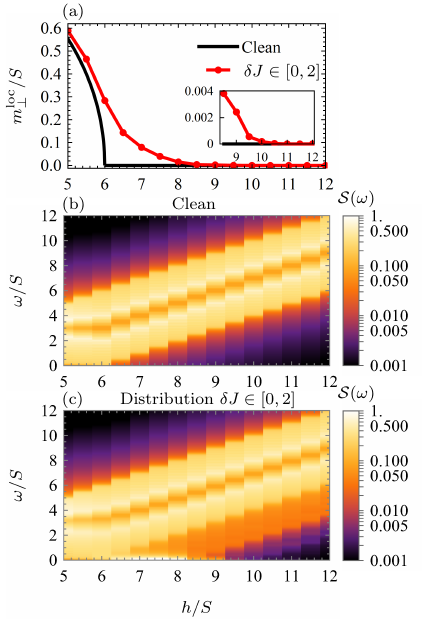}
\caption{
Heisenberg antiferromagnet in a magnetic field, comparing the clean system to that with $p=5\%$ of impurities .
(a) Deviation of local moments from the field direction, $m_\perp^\mathrm{loc}$ Eq.~\eqref{order_parameter}, reveals a smearing of the phase transition due to impurities.
Momentum-integrated dynamical structure factor (b) without impurities and
(c) with impurities $\dJ{}\in[0,2]$ shows a continuous filling of the bulk gap and Goldstone modes over an extended field regime above $\hcrit$, due to disorder.
Lorentzian broadening with $\eta=0.1$ has been used.
}
\label{fig:avHeisenberg_h}
\end{figure}

\begin{figure}[tb]
\includegraphics{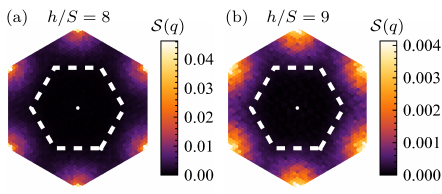}
\caption{
Static structure factors of the Heisenberg antiferromagnet with $\dJ{}\in[0,2]$ and at (a) $h/S=8$ and (b) $h/S=9$, with the bulk contributions at $\Gamma$ and $\Gamma'$ removed. Intensity is concentrated around the $\Gamma'$ points, which is indicative of N\'eel order with broken U(1) symmetry, and originates primarily from isolated impurities in (a) and from impurity clusters exclusively in (b).
}
\label{fig:avHeisenberg_ssf}
\end{figure}

\begin{figure}[tb]
\includegraphics{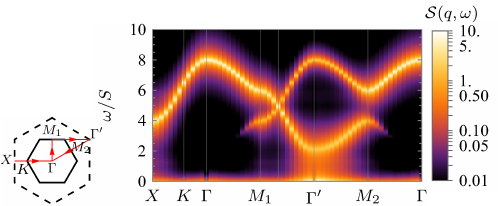}
\caption{
Dynamic structure factor for $h/S=8$ and an impurity distribution $\dJ{}\in[0,2]$ over a path in momentum space. Impurities fill the bulk gap and induce Goldstone modes, the latter due to U(1) symmetry breaking.
Lorentzian broadening with $\eta=0.1$ has been used.
}
\label{fig:avHeisenberg_dsf}
\end{figure}


\begin{figure*}[tb]
\includegraphics{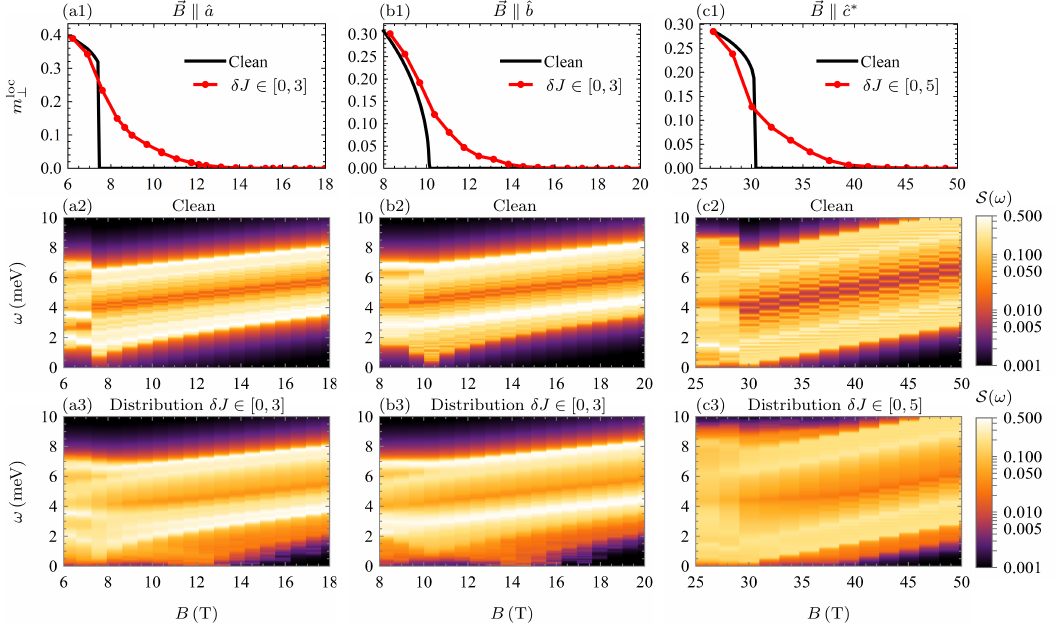}
\caption{As in Fig.~\ref{fig:avHeisenberg_h}, but for the \hkg\ model with a distribution of Heisenberg impurities and magnetic fields in three high-symmetry directions.
(a1-c1)
The phase transition between the canted zigzag and polarized phases is smeared by disorder.
(a2-c3)
Disorder gives rise to low-energy states above $\Bcrit$, which are especially prominent for in-plane fields.
}
\label{fig:dJ_fieldEvolution}
\end{figure*}

\begin{figure*}
\includegraphics{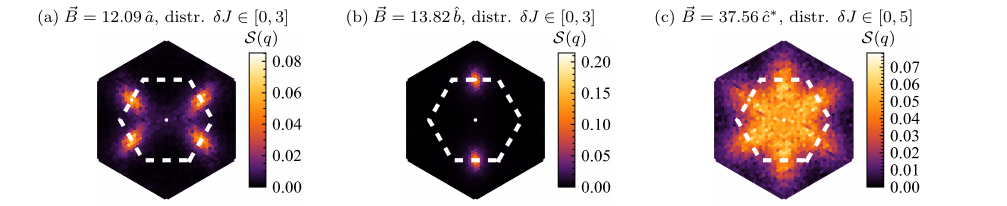}
\caption{Static structure factors for the {\hkg} model with Heisenberg impurities show different types of zigzag order induced by impurities above the bulk critical field. Magnetic fields are in units of T and impurity strengths in meV.
}
\label{fig:dJ_structureFactorsStatic}
\end{figure*}

\begin{figure*}
\includegraphics{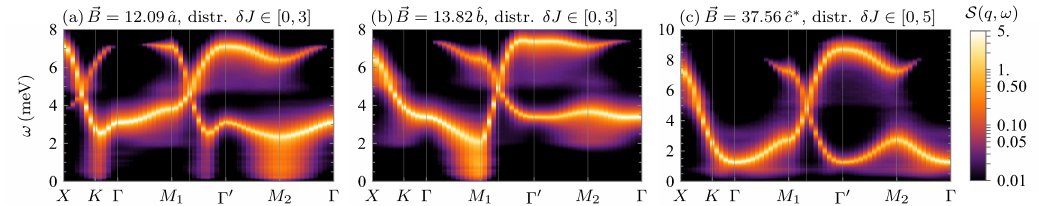}
\caption{Dynamic structure factors for the {\hkg} model with Heisenberg impurities.
Low-energy excitations continuously fill the bulk gap in the case of in-plane fields.
The path in momentum space is as shown in Fig.~\ref{fig:avHeisenberg_dsf}.
Lorentzian broadening with $\eta=0.1$ has been used, magnetic fields are in units of T and impurity strengths in meV.
}
\label{fig:dJ_structureFactorsDynamic}
\end{figure*}


\begin{figure*}[htb]
\includegraphics{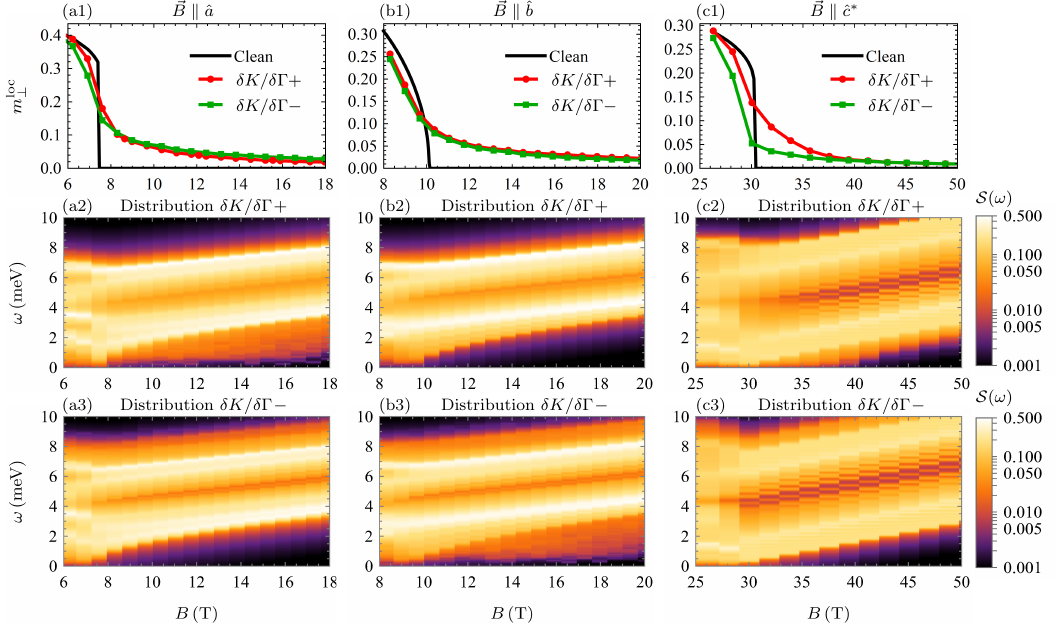}
\caption{
As in Fig.~\ref{fig:dJ_fieldEvolution}, but for two distributions of Kitaev and $\Gamma$ impurities (see text in Sec.~\ref{sec:manyimp_KGdistr} for definition of distributions).
(a1-c1) The phase transition is now replaced by a crossover for both impurity distributions, with full polarization only reached for $B\rightarrow \infty$.
(a2-c3) The presence of in-gap states strongly depends on the combination of field direction and disorder distribution.
}
\label{fig:KG_fieldEvolution}
\end{figure*}

\begin{figure*}[bt]
\includegraphics{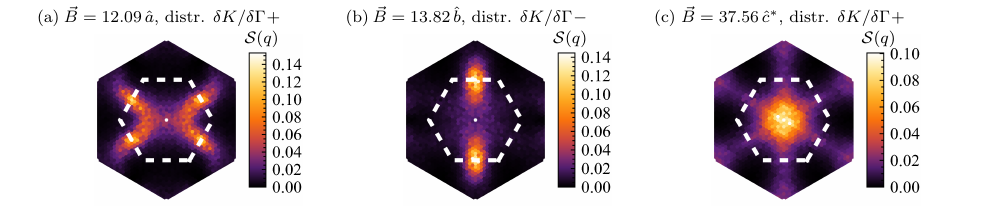}
\caption{
As in Fig.~\ref{fig:dJ_structureFactorsStatic}, but for two distributions of Kitaev and $\Gamma$ impurities (see text for definition).
}
\label{fig:KG_structureFactorsStatic}
\end{figure*}

\begin{figure*}[bt]
\includegraphics{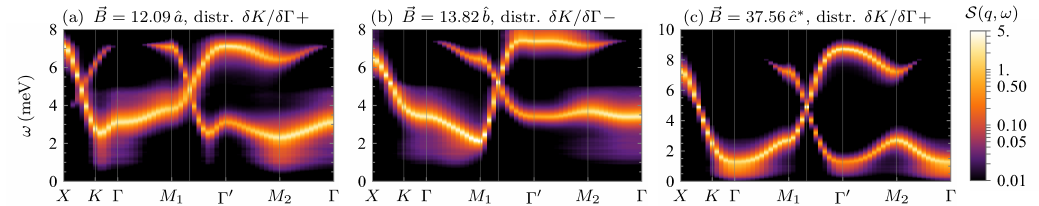}
\caption{
As in Fig.~\ref{fig:dJ_structureFactorsDynamic}, but for two distributions of Kitaev and $\Gamma$ impurities (see text for definition).
While low-energy excitations exist, they exhibit a significant gap due to the absence of symmetry breaking.
}
\label{fig:KG_structureFactorsDynamic}
\end{figure*}


\section{Finite impurity concentration}
\label{sec:manyimp}

In order to connect to experiments, we now turn to study finite impurity concentrations. We will consider different distributions of disorder, to be discussed below, and compute the dynamic spin structure factor in order to inspect the impurity-induced low-energy states. We also discuss the fate of the thermodynamic bulk phase transition upon introducing impurities.

The numerical results shown below are for a defect concentration of $p=5\%$, systems of linear size $L=24$, and disorder averages taken over 40 realizations, unless noted otherwise. With $p=5\%$, effects of isolated impurities, which scale linearly with $p$, dominate most quantities, but higher-order effects are visible as well. While experimentally relevant impurity concentrations may be significantly smaller, their effects would be more difficult to observe in our numerics due to limited system size.

In the following discussion, we will use the term ``ordered'' to refer to the magnetic states displaying symmetry breaking. This can imply either a state with long-range order or a spin glass with short-range order; transitions between long-range order and spin glass as function of field or impurity concentration can occur as well. Our finite-size scaling analyses of the static structure factor at fields above the bulk transition have remained inconclusive in that respect.


\subsection{Disorder distributions and their symmetries}
\label{sec:manyimp_distributions}

A finite concentration of randomly distributed impurities generally reduces the symmetries obeyed by the Hamiltonian as compared to the case of a single impurity, discussed in Sec.~\ref{sec:1imp_symmetries}. Inversion symmetry \inversion{0} is broken in all cases, while randomly distributing impurities on the $x$, $y$ and $z$ bonds in the \hkg\ model may restrict symmetries even further. As seen in the middle section of Table \ref{table:symmetries}, the symmetries allowed by isolated Heisenberg impurities do not depend on the orientation of the bond defect and may, therefore, continue to hold. On the other hand, the orientation of $\dK{}$ and $\dG{}$ defects strongly affects the symmetries present. In our particular model, these symmetries are contradictory to each other and, therefore, all symmetries are explicitly broken. This is summarized in the bottom section of Table \ref{table:symmetries}.

Consequently, we shall study two types of impurity distributions, namely
(i) Heisenberg impurities only, such that residual symmetries may still be broken spontaneously, and
(ii) simultaneous Kitaev and $\Gamma$ impurities for which all symmetries are broken and which therefore represent the more generic situation.

\subsection{Heisenberg limit}
\label{sec:manyimp:Heisenberg}

Before turning to the {\hkg} model, we again consider the Heisenberg antiferromagnet in a magnetic field. Importantly, the model retains its higher symmetry, compared to the \hkg\ model, even in the presence of a finite concentration of bond impurities, since the U(1) spin rotation symmetry about the field axis remains intact.
As before, we choose $J=1$ as energy unit in this subsection. Numerical results have been obtained for $p=5\%$ defects, with the defect strength drawn from a box distribution $\dJ{}\in [0,2]$.

Fig.~\ref{fig:avHeisenberg_h} shows an overview of impurity-induced phenomena and their field evolution.
The fate of the phase transition between the canted antiferromagnetic and polarized phases can be assessed from $m_\perp^\mathrm{loc}$, as defined in Eq.~\eqref{order_parameter} and shown in Fig.~\ref{fig:avHeisenberg_h}(a). In contrast to the sharp transition that takes place in the clean case, one sees a gradual onset of $m_\perp^\mathrm{loc}$ upon lowering the field in the presence of disorder. This is indicative of a smeared phase transition.
Spin-wave excitations are revealed in the momentum-integrated dynamical structure factor, Fig.~\ref{fig:avHeisenberg_h}(b,c). Prominent impurity-induced zero modes are seen in Fig.~\ref{fig:avHeisenberg_h}(c) in an extended field region above $\hcrit$. In contrast to the single-impurity case where they are well-separated from the bulk spectrum, Fig.~\ref{fig:Heisenberg}(a), here the clean-limit gap is continuously filled.

To discuss the smearing of the phase transition, we first focus on the dilute limit: Using the results of Sec.~\ref{sec:1imp_Heisenberg}, we can define $\hcrit^{\rm dil}=\hcrit^\ast(dJ=dJ_{\rm max})$, corresponding to the field where the strongest impurity, if isolated, induces a texture upon lowering the field. The continuous distribution of impurity strengths then leads to the onset of local textures at different fields below $\hcrit^{\rm dil}$, leading to a smearing of the transition. For the specific parameters, $\hcrit^{\rm dil}/S = 8.5$.
However, from Fig.~\ref{fig:avHeisenberg_h}(a) we deduce weak but non-vanishing order also at higher fields. This represents physics beyond the dilute limit: Any given disorder realization involves (rare) impurity clusters, i.e., defects in close spatial vicinity. While their effect is weak, they stabilize a textured phase above $\hcrit^{\rm dil}$, since an impurity cluster effectively acts as a stronger impurity. This is nicely seen in the static structure factor in Fig.~\ref{fig:avHeisenberg_ssf}, where panels (a) and (b) are for $h<\hcrit^{\rm dil}$ and $h>\hcrit^{\rm dil}$, respectively. In both cases, the weight around $\Gamma'$ indicates inhomogeneous canted N\'eel order, with much lower intensity and shorter correlation length in (b) compared to (a).
Taking rare events into account, our bounded distribution yields a true transition field $\hcrit^{\ast\ast}/S=18$, corresponding to a large region of bonds of strength $J+\dJ{}=3$. However, due to the extreme rarity of such events, $m_\perp^\mathrm{loc}$ develops gradually at lower fields.

Goldstone modes in the spin-wave spectrum are a direct result of the broken U(1) symmetry at $h<\hcrit^{\ast\ast}$. The role of $\hcrit^{\rm dil}$ is manifest in the momentum-integrated structure factor, Fig.~\ref{fig:avHeisenberg_h}(c), where weight at lowest energies is significant for $\hcrit<h<\hcrit^{\rm dil}$ but gets much weaker for $\hcrit^{\rm dil}<h<\hcrit^{\ast\ast}$. The former receives contributions from individual impurities and scales linearly with $p$, while the latter arise from impurity clusters and scales with higher powers of $p$. A true gap opens for $h>\hcrit^{\ast\ast}$.

The momentum-resolved dynamic structure factor, Fig.~\ref{fig:avHeisenberg_dsf}, reveals more details about the excitations. While the bulk modes display a clear gap (and the Dirac points expected for the honeycomb lattice), the continuous impurity distribution induces finite-energy in-gap modes which are concentrated around the ordering wavevector $\Gamma'$. Interestingly, the modes at very low energy are completely smeared in momentum space, indicating that the order parameter distribution, and with it the Goldstone-mode weight, is spatially extremely inhomogeneous.


\subsection{Distributions of Heisenberg impurities}
\label{sec:manyimp_Jdistr}

Distributions of Heisenberg impurities (as opposed to generic impurities) in the {\hkg} model allow for some symmetries of the Hamiltonian to remain, as shown in Table~\ref{table:symmetries}. Moreover, all Heisenberg impurities belong to type II, which implies that the polarized, i.e., non-textured, classical state is preserved at high magnetic fields. Not unlike the case of the Heisenberg antiferromagnet discussed above, symmetries can be broken at fields above the bulk $\Bcrit$, accompanied by low-energy in-gap states. In order to observe this phenomenon, we use the insights of the single-impurity analysis, Secs.~\ref{sec:1imp:broken} and \ref{sec:1imp_KG}, to select appropriate disorder distributions. We choose a box distribution of $\dJ{} \in [0,3]$ for $\ha$ and $\hb$ and $\dJ{} \in [0,5]$ for $\hc$ with a defect concentration of 5\% in all cases. As before, we will quote energies in meV and magnetic fields in T.

Similar to the case of the Heisenberg antiferromagnet, the phase transition between the symmetric high-field phase and the symmetry-broken zigzag phase is smeared, as seen in Fig.~\ref{fig:dJ_fieldEvolution}(a1-c1).
Furthermore, there exist impurity-induced low-energy magnetic states with significant weight below the bulk gap above $\Bcrit$, as revealed in the momentum-integrated dynamic structure factor, Fig.~\ref{fig:dJ_fieldEvolution}(a2-c3).

As with the Heisenberg antiferromagnet, the phenomena contributing to smearing the transition are the successive onset of textures due to the continuous distribution of impurities strengths, from isolated impurities and from impurity clusters at higher magnetic fields.
While isolated impurities enabled different textures to occur depending on the placement of the defect bond, Fig.~\ref{fig:SSFbroken}, for a finite impurity concentration we observe canted zigzag order with an orientation determined by the field direction, Fig.~\ref{fig:dJ_structureFactorsStatic}.
Interestingly, the cases $\ha$ and $\hc$ lead to classical states consisting of two and three simultaneous zigzag orientations respectively, each induced by impurities on one of the three types of bonds, a phenomenon that is not present in the clean system.

As before, we can define $\Bcrit^{\rm dil}$ corresponding to the field below which the strongest impurity, if isolated, induces a texture. For the parameters in Fig.~\ref{fig:dJ_fieldEvolution}, the values of $\Bcrit^{\rm dil}$ are 12.6~T, 14.7~T, and 39.5~T in panels (a,b,c), respectively. Similar to the Heisenberg antiferromagnet, the impurity-induced order parameter is finite but very small above $\Bcrit^{\rm dil}$, as it is exclusively carried by impurity clusters. The true transition to the high-field phase happens at a much larger $\Bcrit^{\ast\ast}$ which is, e.g., about 30~T for $\hb$.

Significant in-gap states that approach zero energy are observed for in-plane fields near $\Bcrit^\mathrm{dil}$, Fig.~\ref{fig:dJ_fieldEvolution}(a3,b3). Details can be seen in the momentum-resolved structure factor in Figs.~\ref{fig:dJ_structureFactorsDynamic}(a,b), where the gap is filled at the locations in momentum space where the minima of the bulk spectrum are found. These are the ordering wave vectors $M$ and, in the case of $\ha$, $K$.
The low energy of these modes is a remnant of the single-impurity phase transition, where the in-gap modes of strong isolated impurities tend to soften. Since the true transition only happens at higher fields, however, these are not true zero modes.

For fields $B>\Bcrit^\mathrm{dil}$, the in-gap modes originating from isolated impurities develop a gap, and excitations from impurity clusters become visible below them, Fig.~\ref{fig:dJ_fieldEvolution}(a3,b3). For $B<\Bcrit^\mathrm{dil}$ the spectrum develops a small apparent gap; given the finite-size limitations of our simulations we are not able to quantify it. This reflects the fact that there are no zero modes from isolated impurities close to $\Bcrit$, as seen in Fig.~\ref{fig:KG_comb}(a1,b1).

In contrast, for $\hc$ impurity-induced excitations form a continuum below the bulk gap, but do not soften anywhere above $\Bcrit$, Figs.~\ref{fig:dJ_fieldEvolution}(c3). Instead, a small gap is always visible, as in Fig.~\ref{fig:dJ_structureFactorsDynamic}(c). This is because here the symmetry-breaking texture of an isolated impurity forms via a first-order transition with finite-energy excitations at the transition, Fig.~\ref{fig:broken}(c1-c3). We expect this to change for a wider impurity distribution, since the first-order transition becomes continuous for stronger impurities, Fig.~\ref{fig:KG_comb}(c1).


\subsection{Distributions of Kitaev and $\Gamma$ impurities}
\label{sec:manyimp_KGdistr}

\renewcommand{\labelenumi}{(\roman{enumi})}

As discussed in Sec.~\ref{sec:manyimp_distributions}, distributions of Kitaev and $\Gamma$ defects explicitly break all symmetries of the \hkg\ model, see Table~\ref{table:symmetries}, with consequences both on the ground state and the spin-wave excitations.
Since we target situations with impurity-induced low-energy states, we utilize the single-impurity results, Sec.~\ref{sec:1imp_KG}. As shown in Fig.~\ref{fig:KG_comb}, in-gap states require moderate impurity strengths when $\dK{}$ and $\dG{}$ are combined on the same bond. Therefore, we choose the following impurity distributions:
\begin{enumerate}
\item
$\dK{}\!\in\! [0,5]$ and $\dG{}\!\in\![0,2.5]$, which we will call \kgp, where $K$ becomes weaker and $\Gamma$ stronger, and
\item
$\dK{}\!\in\![-5,0]$, $\dG{}\!\in\![-2.5,0]$, which we call \kgm, where $K$ becomes stronger and $\Gamma$ weaker.
\end{enumerate}
In both cases we leave $\dK{}$ and $\dG{}$ uncorrelated.

The absence of any symmetries results in the phase transition between the canted zigzag and the high-field phases disappearing entirely, regardless the impurity distribution, as seen in Fig.~\ref{fig:KG_fieldEvolution}(a1-c1). A homogeneous polarized state is approached only as $B\rightarrow\infty$ and, thus, phase transitions are replaced by crossovers.
Moreover, the existence of in-gap states strongly depends on the direction of the magnetic field for each impurity distribution, Fig.~\ref{fig:KG_fieldEvolution}(a2-c3). This is a result of the strong spin-orbit coupling encoded in the $K$ and $\Gamma$ terms of the Hamiltonian that now carry disorder. In particular, the \kgp\ distribution induces significant in-gap states for $\ha$ and, to a lesser extent, for $\hc$, while the \kgm\; distribution does so for $\hb$.

The vanishing of the phase transition can be interpreted by use of the single-impurity analysis, Sec.~\ref{sec:1imp:unbroken}: distributions include impurities of type I, as defined in Sec.~\ref{sec:1imp_symmetries}, which always induce textures regardless of impurity or field strength. As a result, the disordered system remains in an inhomogeneous textured state, even at high fields. For very weak disorder, however, we expect that first-order transitions without symmetry breaking would remain intact.

Such textured states are described by the static structure factors in Fig.~\ref{fig:KG_structureFactorsStatic}(a-c). They feature zizag-type spin correlations similar to the clean-limit state below $\Bcrit$, not unlike those in Fig.~\ref{fig:dJ_structureFactorsStatic}. However, for $\ha$, double-$Q$ states occur in the vicinity of some impurities, which are characterized by Fig.~\ref{fig:SSFunbroken}(a) in the single-impurity limit.
The structure factors also indicate progressive weight accumulation around (but not at) $\Gamma$, particularly pronounced in Fig.~\ref{fig:KG_structureFactorsStatic}(c). This broad peak at $\Gamma$ simply corresponds to an almost polarized, but inhomogeneously textured, state.

In-gap states, in the cases when they exist, can be seen in more detail in the momentum-resolved dynamic structure factors, Fig.~\ref{fig:KG_structureFactorsDynamic}.
Even though low-energy states dominate the majority of the bulk gap, they nevertheless leave a distinct albeit small gap, also seen in Fig.~\ref{fig:KG_fieldEvolution}(a2,b3,c2). This is to be expected, since in the case of isolated impurities gap closings signified symmetry breaking. In distributions studied in this section, however, there are no symmetries to be broken.


\section{Experimental relevance of impurities for $\alpha$-R\lowercase{u}C\lowercase{l}$_3$}

\begin{figure*}
\includegraphics{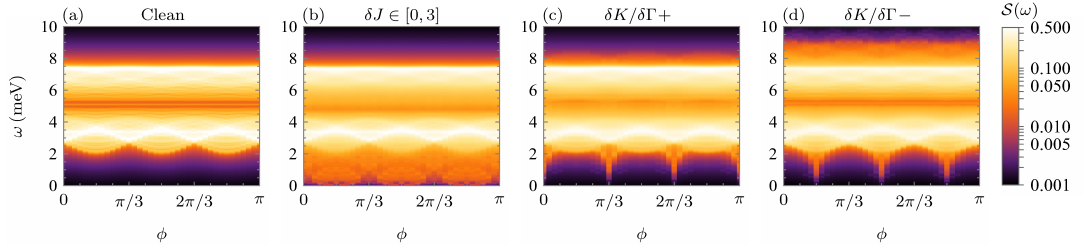}
\caption{Momentum-integrated dynamical structure factor at fixed magnetic field strength $B=13.82$~T as function of the in-plane field direction $\phi$, with $0$ corresponding to $\ha$. While the filling of the bulk gap is mostly isotropic for a distribution of Heisenberg impurities (b), it shows strong direction-dependence for distributions of $K$ and $\Gamma$ impurities (c,d).
}
\label{fig:rotation}
\end{figure*}

We now turn to connect our results for the {\hkg} model to the van-der-Waals material {\rucl}. As mentioned in the introduction, several experiments point to the presence of low-energy magnetic excitations in a field regime far above the bulk critical field, e.g., at 15~T. Given that such excitations are not expected in the clean limit, we consider them to be the identifying signature of bond disorder. Concretely, the field dependence of the thermal conductivity at very low temperature \cite{hentrich2020} implies the existence of low-energy magnetic excitations which scatter phonons. Furthermore, in NMR measurements \cite{baek2017,baek2020}, the exponential temperature decay of the relaxation time $T_1$, which is characteristic of a gapped spectrum, is cut off at low temperature turning into a plateau of $T_1$ indicating a filling of the gap. Empirical modeling suggests a broad distribution of in-gap energies, reminiscent of an impurity band.

We note that other experiments have thus far not detected low-energy in-gap states at elevated fields, most notably ESR \cite{ponomaryov2017}, Raman spectroscopy \cite{wulferding2020}, and inelastic neutron scattering \cite{balz2019}. We see two reasons: First, for small impurity concentrations the signal will be weak and difficult to distinguish from background contributions. Second, the strongest intensities can be expected near momenta K and M, Figs.~\ref{fig:dJ_structureFactorsDynamic} and \ref{fig:KG_structureFactorsDynamic}; these cannot be directly probed by ESR and Raman, and they have not been accessed in the neutron experiment of Ref.~\onlinecite{balz2019}.

\subsection{Disorder distributions and experimental signatures}

In the following, we use the numerical results of Sec.~\ref{sec:manyimp} to draw conclusions concerning the likely type of (dominant) disorder in \rucl. Given that the phenomena arising from $K$ and $\Gamma$ impurities strongly depend on the direction of the applied field, a comparison between experiment and theory requires the knowledge of the field direction. In the aforementioned experiments \cite{hentrich2020,baek2017,baek2020} the field was applied in-plane but, unfortunately, the precise direction was not specified. This uncertainty leaves us with two distinct scenarios, that we outline in the following.

(i) Disorder may be dominated by strong Heisenberg impurities. In this case, the presence of impurity-induced low-energy states does not depend on field direction, owing to the isotropic nature of the impurity. This is illustrated in Fig.~\ref{fig:rotation}(b), where the continuum of in-gap states shows only minor shifts upon field rotation, on par with the bulk spectrum.
This type of impurities, however, ought to be strong in comparison to the bulk Heisenberg coupling. This may be realized due to the fact that the clean Hamiltonian is a result of cancellation of isotropic interactions, sensitive to bond angles.

(ii) On the other hand, disorder may be dominated by moderate Kitaev and $\Gamma$ impurities. In this case, the presence of in-gap states depends strongly on the direction of the magnetic field, which can be understood by the fact that these impurities encode information about strong spin-orbit coupling. This is seen in Figs.~\ref{fig:rotation}(c,d), where two different impurity distributions induce low-energy states only at fields near a (different) high-symmetry direction.

Therefore, we suggest experimental set-ups that can detect low-energy excitations be performed with a systematic analysis of directions of magnetic field. Then, information about the true impurity distributions in {\rucl} can be obtained.

It must be pointed out that even though calculations were performed for a model with specific parameters and impurity concentrations of $p=5\%$, all impurity-induced phenomena can be explained by symmetry arguments. Therefore, we do not expect our main results to be sensitive to parameter choice and different impurity concentrations would only change the weight of impurity-induced states.

\subsection{Quantum corrections and finite temperatures}

Results in this paper have been computed to leading order is $1/S$ and at zero temperature. However, quantum corrections and finite temperatures are highly relevant to any experimental set-up.

Quantum corrections to our results can be computed systematically in a $1/S$ expansion \cite{consoli2020}, while results for very small systems can alternatively be obtained by exact diagonalization directly for $S=1/2$. Quantum corrections will modify the numerical values for phase boundaries, moment amplitudes, and canting angles. They will, however, not change any of the qualitative conclusions of this paper, which are based on symmetries, provided that the clean-limit zero-field model is located in the regime of robust zigzag order.

At finite temperatures, our key feature of interest, the impurity-induced in-gap states, continue to be present and well defined, as long as $T$ is significantly smaller than the clean bulk gap. In contrast, the weakly ordered states above the bulk critical field will have a correspondingly low ordering temperature \cite{order_foot}, and hence their detection would require thermodynamic measurements at extremely low $T$.

It is interesting to note that, for generic impurities, the finite-field canted zigzag state does not break any symmetries of the Hamiltonian, see Sec.~\ref{sec:manyimp_KGdistr}. This implies that the finite-temperature transition into this state, if originally continuous, will no longer be sharp, but become a crossover due to disorder effects. The transition remains well defined; however, at zero field where time reversal is spontaneously broken by the zigzag state. Indeed, progressive smearing of the zigzag transition with increasing field has been seen experimentally \cite{wolter2017}.


\section{Summary}
\label{sec:summ}

Motivated by recent experiments \cite{baek2017,baek2020,hentrich2020} that revealed signatures of quenched disorder in nominally clean samples of \rucl, we studied the role of bond defects in extended {\hkg} models, with a focus on parameters relevant for \rucl. In particular, we examined the role of defects in the asymptotic high-field phase in the low-temperature limit. We have shown that dilute impurities can generate both in-gap states and magnetic textures in their vicinity; because of the low symmetry of the system the latter can either be accompanied by spontaneous symmetry breaking or can respect all symmetries. A finite concentration of impurities leads to in-gap states forming an impurity band, and the presence of textures either smears the bulk transition to the high-field phase, or destroys it in favor of a crossover. The details of the impurity-induced low-energy states, however, depend sensitively on the type of impurities and on the field direction. This analysis has enabled us to propose two distinct scenarios for the character of bond disorder in {\rucl}, and we make concrete predictions for experiments which could discriminate between the two scenarios.

While our study was motivated by experiments on {\rucl}, with calculations performed for a specific parameter set, our general analysis and symmetry arguments can be adapted to any antiferromagnet in a magnetic field.  More broadly, our work highlights the relevance of impurity effects in spin-orbit-coupled magnets, and we hope that this study will help highlight the consequences of disorder in the hunt of Kitaev materials.


\acknowledgments

We thank E. C. Andrade, C. Hess, L. Janssen, F. K\"ohler, and A. Wolter for illuminating discussions and collaboration on related work.
This work was funded by the Deutsche Forschungsgemeinschaft (DFG) through SFB 1143 (project id 247310070) and the W\"urzburg-Dresden Cluster of Excellence on Complexity and Topology in Quantum Matter -- \textit{ct.qmat} (EXC 2147, project id 390858490).

\bigskip

\textit{Note added --} We recently became aware of Ref.~\onlinecite{andrade24} which also studies defects in Heisenberg-Kitaev models. The authors also discuss impurity-induced in-gap states, with results broadly consistent with ours, but the focus of Ref.~\onlinecite{andrade24} is on the detection of impurity states via scanning tunneling spectroscopy.


\appendix

\renewcommand{\dbltopfraction}{1}
\begin{figure*}
\includegraphics[scale=1]{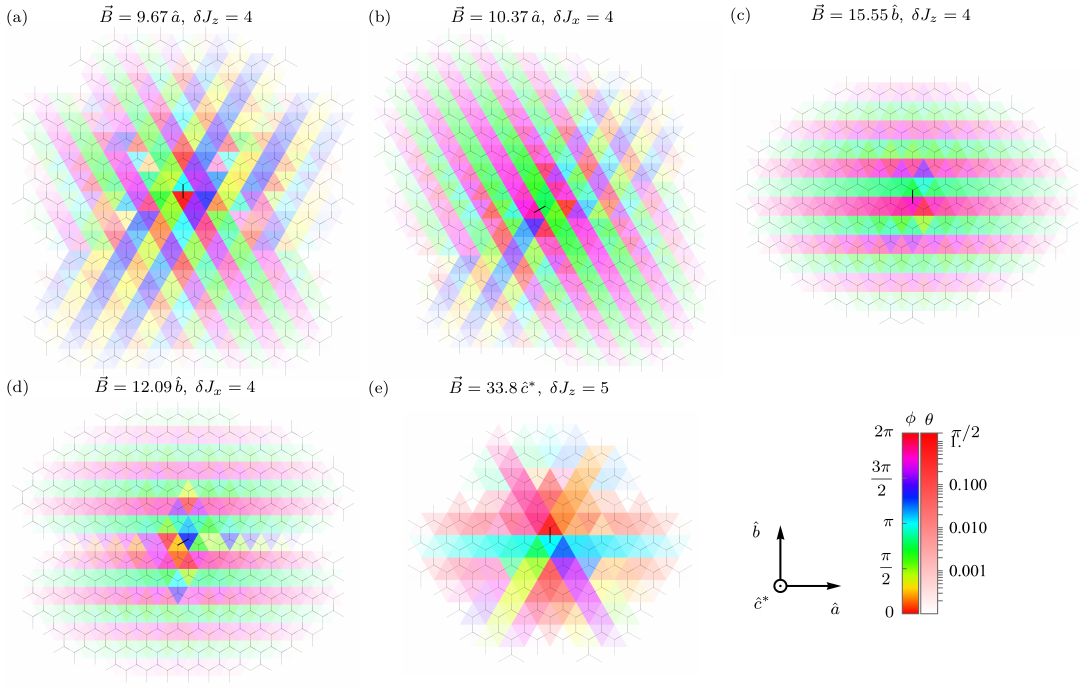}
\caption{
Top-down view of symmetry-breaking textures for all distinct combinations of field direction and impurity bond; the corresponding spin structure factors are in Fig.~\ref{fig:SSFbroken}. Each spin is represented by a triangle, with its color representing the spin direction in the plane perpendicular to $\vec{B}$ and its brightness the angle relative to $\vec{B}$. The figures indicate combinations of different zigzag patterns which decay with distance from the impurity.
Magnetic fields are quoted in T and impurity strengths in meV.
}
\label{fig:textures2d_broken}
\end{figure*}

\begin{figure*}
\includegraphics[scale=1]{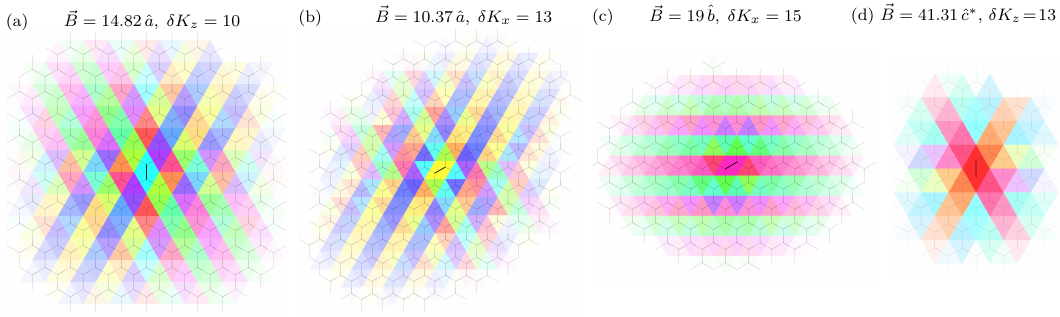}
\caption{Top-down view of symmetric textures with color scheme as in Fig.~\ref{fig:textures2d_broken}. Spin configurations preserve \reflection{z} and \inversion{0} in (a,d) and just \inversion{0} in (b,c). The corresponding spin structure factors are in Fig.~\ref{fig:SSFunbroken}.
}
\label{fig:textures2d_unbroken}
\end{figure*}

\section{Texture description}
\label{app:textures}

As discussed in Secs.~\ref{sec:1imp:broken} and \ref{sec:1imp:unbroken}, impurities can induce textures with or without symmetry breaking. The static structure factors, Figs.~\ref{fig:SSFbroken} and \ref{fig:SSFunbroken}, reveal that textures differ greatly depending on the directions of the magnetic field and the defect bond. Here we analyze the real-space configuration of textures for all distinct combinations of field and impurity bond.

\subsection{Symmetry-breaking textures}
\label{app:textures_broken}

In Fig.~\ref{fig:textures2d_broken} we show a representation of symmetry-breaking textures. The deviation $\theta$ of spins from the direction of the magnetic field shows a decay with distance from the center of the impurity. At the same time the spin projections transverse to the magnetic field form distinct patterns described by the angle $\phi$.
Since these textures are formed via a spontaneous symmetry-breaking mechanism, they act as precursors to the low-field phase. Therefore, the pattern of each texture is determined by a combination of the low field order present for each field direction and a state minimizing the energy of the defect bond.

We start by analyzing textures in a magnetic field $\hb$, as it only allows for a unique zigzag orientation with ferromagnetic chains that are perpendicular to the field.
An antiferromagnetic $\dJ{z}$ impurity is positioned perpendicularly to the favored zigzag orientation. Therefore, the resulting texture shows a clear canted zigzag pattern that is transverse to the magnetic field (Fig.~\ref{fig:textures2d_broken}(c)). On the other hand, an antiferromagnetic $\dJ{x}$ is incompatible with the state favored by the magnetic field. In this case, the texture's spin configuration changes character as function of the distance from the defect (Fig.~\ref{fig:textures2d_broken}(d)). While at large distances we observe the same zigzag orientation perpendicular to the field, this is greatly distorted closer to the impurity in order to accommodate the defect bond.
The short-distance behavior is reflected in the static structure factor Fig.~\ref{fig:SSFbroken}(d), showing the remnants of a zigzag pattern perpendicular to the impurity bond.

A magnetic field $\ha$ allows for two degenerate zigzag orientations. An antiferromagnetic $\dJ{x}$ impurity locally uniquely selects the one perpendicular to the $x$ bond as seen in Fig.~\ref{fig:textures2d_broken}(b). However, at larger distances the texture is split into domains with different zigzag orientations.
In Fig.~\ref{fig:textures2d_broken}(a), we can see the texture that results from $\dJ{z}>0$ impurity, which is incompatible with either of the zigzag directions favored by the magnetic field. As a result, the different zigzag domains that are prevalent are larger distances merge into a new unique pattern.

The case when $\hc$ all three zigzag orientations are degenerate and
a large number of symmetries break in the formation of a texture. The resulting texture is seen in Fig.~\ref{fig:textures2d_broken}(e), featuring zigzag ferromagnetic chains in all three directions. However, the one perpendicular to the defect bond is dominant, which can also be deduced from the static structure factor in Fig.~\ref{fig:SSFbroken}(e).

\subsection{Symmetric textures}
\label{app:textures_unbroken}

Textures forming without symmetry breaking must preserve the residual symmetries while they also have to conform to the types of order allowed by the direction of the magnetic field in each case.

An impurity on a z-bond preserves \reflection{z} and inversion when the magnetic field is either in the $\hat{a}$ or $\hat{c}^*$ direction. In both cases, the texture emerging at high fields will be symmetric under both symmetry transformations. However, the type of order as revealed by the static structure factor will differ.
For $\ha$ the two zigzag directions favored by the magnetic field combine into a region with double-Q order around the impurity, Fig.~\ref{fig:textures2d_unbroken}(a).
The symmetric texture for $\hc$ is shown in Fig.~\ref{fig:textures2d_unbroken}(d). Its dominant feature are the two ferromagnetic zigzag chains that intersect at the impurity.

Inversion is the only symmetry present when the impurity is placed on an x- or y-bond with a magnetic field in either of the two in-plane directions.
For $\ha$, \inversion{0} symmetry forces the zigzag with ferromagnetic chains run through the defect bond to become dominant as seen in Fig.~\ref{fig:textures2d_unbroken}(b).
For $\hb$ the \inversion{0}-symmetric texture, Fig.~\ref{fig:textures2d_unbroken}(c), has a simple zigzag order perpendicular to the magnetic field.

\section{Texture decay}
\label{app:decay}

\renewcommand{\dbltopfraction}{0.8}
\begin{figure}[tb]
\includegraphics[scale=1.]{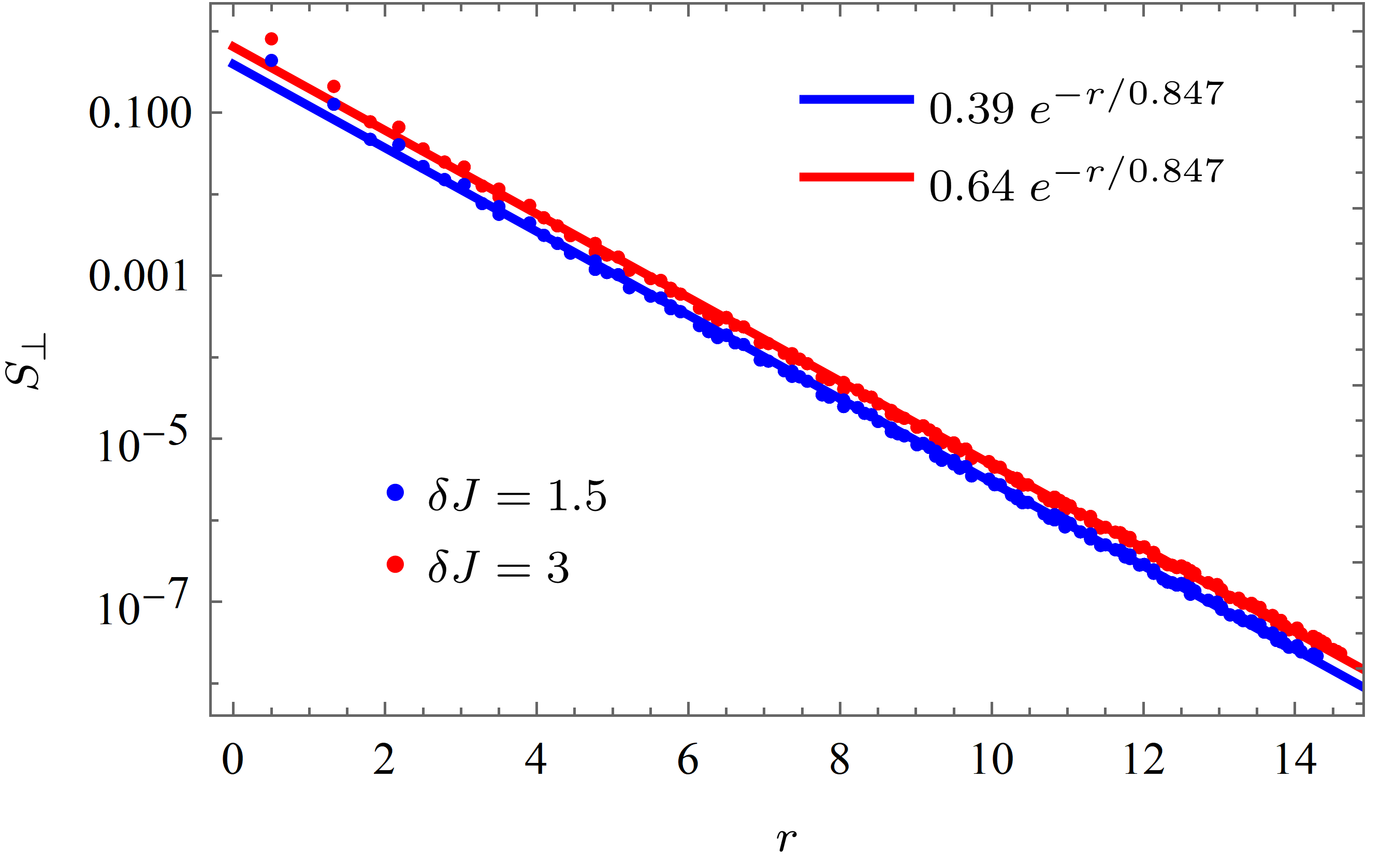}
\caption{Decay of two textures in the Heisenberg antiferromagnet in a field $h/S=7$. $S_\perp$ is the spin projection perpendicular to the field and $r$ is measured in units of bond length.}
\label{fig:decay_Heisenberg}
\end{figure}

\begin{figure}[tb]
\includegraphics{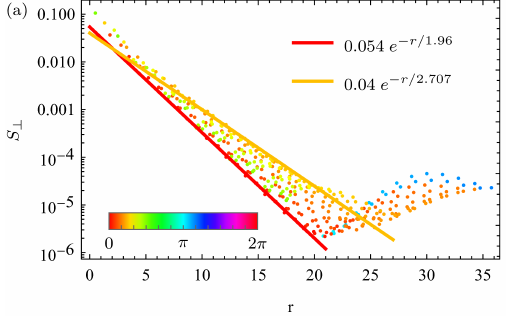}
\includegraphics{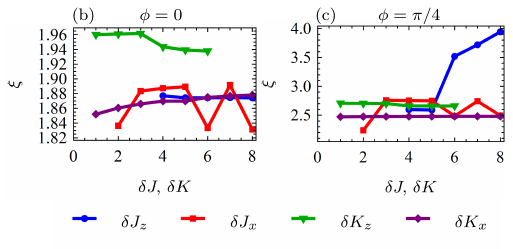}
\caption{(a)
Decay of symmetric textures for $\dK{z}=1$ and $\vec{B}=8.64\,\mathrm{T}\, \hat{a}$. Points of different colors represent lattice sites at different in-plane directions away from the impurity. Solid lines are exponential fits of data selected for $\phi=0$ ($\hat{a}$ direction) in red and $\phi=\pi/4$ in yellow. The scatter at larger distance is an artifact of periodic boundary conditions.
(b,c)
Decay length $\xi$ of textures from different impurities at $B=8.64$~T along two in-plane directions.
}
\label{fig:decay_HKG}
\end{figure}

In the case of isolated impurities, textures are a result of the bulk's response to a local perturbation.
For small perturbations, linear-response theory dictates that the decay length of textures must be equal to the bulk correlation length. We will test the validity of linear response by studying the decay of textures arising from different impurities in the same bulk.

In the simple case of impurities in the Heisenberg model, textures decay exponentially in an isotropic way as
\begin{equation}
S_\perp(r)=c\, e^{-r/\xi},
\end{equation}
where $S_\perp$ is the projection of each spin on the plane perpendicular to the magnetic field, $r$ measures the distance form the center of the impurity bond and $\xi$ is the decay length.
In Fig.~\ref{fig:decay_Heisenberg} we present the decay of two textures originating from different impurity strengths at the same field value. While the prefactor $c$ of the decay depends on the impurity strength, the two textures have almost identical decay lengths.
Therefore, we conclude that the response to the defect is determined by bulk properties.

Textures in the \hkg\ model, on the other hand, do not decay isotropically. An example for a symmetric texture is shown in Fig.~\ref{fig:decay_HKG}(a), where the broad range of $S_\perp$ at larger distances is clearly demonstrated. The fastest decay takes place in the direction of the magnetic field (red), while the slowest in the direction at a $45^\circ$ angle away from that (yellow).

The decay of different textures has to be compared in order to determine whether bulk behavior is dominated by linear response. While in the Heisenberg case the only free parameter was the impurity strength, in the \hkg\ case we can change the orientation of the defect bond, as well as the type of impurity thus creating symmetric or symmetry-breaking textures. In Figs.~\ref{fig:decay_HKG}(b,c) we compare the decay lengths of four textures along the directions for $\phi=0$ and $\phi=\pi/4$.
In the case of symmetric textures from $\dK{}$ impurities, the decay length remains mostly unaffected by the impurity strength. This can be traced to the fact that any infinitesimally weak impurity can form a texture and we can therefore apply linear-response theory. Interestingly, in Fig.~\ref{fig:decay_HKG}(b) the transition from a symmetric to a symmetry-breaking texture can be seen for $\dK{z}\geq 4$. Textures from impurities on different bonds, however, have slightly different decay lengths, indicating the need for corrections to linear response.
The decay length of symmetry-breaking textures can depend strong on impurity strength. This can be understood due to the fact that there is a minimum $\dJ{}$ required to create a texture for any $B>\Bcrit$. As a result, the limit $\dJ{} \rightarrow 0$ is not valid and, therefore, linear-response theory breaks down.
\\



\begin{thebibliography}{99}
\bibitem{imamura_foot}
Defect-induced spin excitation have also been studied in electron-irradiated {\rucl} in Ref.~\onlinecite{imamura23}, but only for fields up to 12~T which may not be sufficient to open a sizeable bulk spin gap.
\end{thebibliography}
\end{document}